\documentclass[aps,prc,superscriptaddress,showkeys,showpacs,twocolumn,longbibliography,nofootinbib]{revtex4-2}
\usepackage{color}
\usepackage{graphicx}
\usepackage{appendix}
\usepackage{amsmath,amsfonts}
\usepackage{float}
\usepackage[bookmarks,bookmarksnumbered]{hyperref}
\hypersetup{colorlinks = true,
            linkcolor = blue,
            anchorcolor =red,
            citecolor = blue,
            pdfauthor=author}

\newcommand{\nn}{\nonumber}

\graphicspath{{./figures/}}

\begin{document}
\author{Jin Hu}
\affiliation{Department of Physics, Tsinghua University, Beijing 100084, China.}
\author{Shuzhe Shi}\email{shuzhe.shi@stonybrook.edu}
\affiliation{Center for Nuclear Theory, Depart of Physics and Astronomy, Stony Brook University, Stony Brook, NY 11784-3800, USA.}

\title{Multicomponent second-order dissipative relativistic hydrodynamics with binary reactive collisions}
\begin{abstract}
We derive the multicomponent second-order dissipative relativistic hydrodynamic equations using the moment-expansion method. By computing the transport coefficients using hard-sphere interactions, we investigate the role of multiple components and the reactive collisions. We find that both of these factors increase the effective cross-section and hence decrease the transport coefficients and relaxation times. We further compute such transport properties using leading-order perturbative QCD cross-sections. For both types of cross-sections, we find that the ratio between vector current relaxation time and conductivity for a multicomponent fluid is notably different from that for a single-component fluid. Therefore, the current study provides a more applicable guideline for such a ratio in phenomenological hydrodynamics simulations.
\end{abstract}
\maketitle

\section{Introduction}
Relativistic hydrodynamics has been successfully applied to describe the evolution  of the fireball created in relativistic heavy-ion collisions and determine the properties of quantum chromodynamics(QCD) matter by analyzing experimental data produced at BNL-RHIC and CERN-LHC~\cite{Hama:2004rr,Huovinen:2006jp,Jeon:2015dfa,Romatschke:2017ejr,Bernhard:2019bmu,Auvinen:2020mpc,Nijs:2020roc,JETSCAPE:2020mzn,Parkkila:2021tqq}. As part of the continuous effort of searching for the QCD critical point, the physics of high density baryonic matter has long attracted extensive attention and is one of the main tasks at the future FAIR and NICA facilities. Noting that in the collision of heavy nuclei, especially peripheral collisions where the effect of neutron skin starts to engage, the system created could be high in isospin chemical potential, owing to the imbalance between the number of neutrons and protons. The cross effects could be enhanced by large baryon and isospin chemical potential and give rise to observable experimental phenomena. Therefore, a multicomponent dissipative hydrodynamic theory respecting the microscopic property, especially reactions that change the identity of in-coming particles, is called for.

As a low-energy effective theory of large distance and time scale, relativistic hydrodynamics can be derived from the microscopic transport theory --- the Boltzmann equation. The pioneering work constructing causal stable second-order relativistic hydrodynamics was initiated by Israel and Stewart~\cite{Israel:1979wp}, who developed the moment method originally formulated for non-relativistic fluids~\cite{Muller:1967zza,Muller:1999in}. The theory of Israel and Stewart (IS) successfully fixes the acausal problem~\cite{Hiscock:1985zz} long existing in the first-order theory of hydrodynamics and allows a numerically stable implementation, which leads to important applications in modeling relativistic heavy-ion collisions and astrophysics.
 
Nevertheless, the IS theory was originally constructed for a simple system, which equivalently means that the application range is limited to a single component or minor extra effects introduced by multicomponents, for instance, the cross effects originating from interactions between particles of various species. Considering that fluid systems are complex as far as their components are concerned, it is natural to derive the hydrodynamic equations for the case of multicomponent mixture respecting theoretic completeness and consistency. There are pioneer works deriving multicomponent viscous hydrodynamics by extending the IS theory to multicomponent mixtures~\cite{Prakash:1993bt,Monnai:2010qp,Monnai:2010th}, applying the renormalization group method~\cite{Kikuchi:2015swa}, or taking the relaxation time approximation which assumes the same relaxation time for all microscopic processes~\cite{Bhadury:2020ngq}~\footnote{In the preparation of the current manuscript, we note that in a recent, parallel work~\cite{Fotakis:2022usk}, the multicomponent dissipative hydrodynamic theory is derived using the same method. See the note after the Summary and Outlook for a detailed description of similarity and distinction.}.  On the other hand, it has been found that the detailed coupling between the diffusion currents is important in determining the features of multicomponent systems~\cite{Greif:2017byw,Fotakis:2019nbq,Rose:2020sjv,Fotakis:2021diq}. It is crucial to construct the hydrodynamic theory with correct structure of conserved currents evolution and evaluate the transport coefficients respecting the microscopic details.
The momentum expansion method~\cite{Denicol:2012cn} is such a method that starts from the microscopic transport theory, naturally connects the viscous corrections in microscopic and macroscopic states, and henceforth derives the equation of motion for the macroscopic dissipative currents.

In this work, we derive the dissipative multicomponent hydrodynamic equations with binary reactive collisions by employing the momentum expansion method. The results of the hydrodynamic equations are presented in Sec.~\ref{sec.hydro_equations}. Then we further take the hard-sphere potential to systematically study the effect of elastic and inelastic binary collisions between different components (Sec.~\ref{sec.tran_prop.hard_sphere}) and take the leading-order perturbative QCD cross-sections to study the transport properties in a realistic system (Sec.~\ref{sec.tran_prop.pQCD}). After the summarizing in Sec.~\ref{sec.summary}, we further supplement Appendices discussing details about moment expansion method (App.~\ref{app.moment_expansion}), null-modes caused by the matching conditions (App.~\ref{sec.null_matching}), calculation of the evolution matrix (App.~\ref{sec.A_matrix}), details of the hard-sphere (App.~\ref{app.hard_sphere}) and LO pQCD (App.~\ref{sec.pQCD_cross}) interaction integrals.

The conventions and notations in this paper are listed as follows.
\begin{itemize}
\item 
    Superscript/subscripts: 
    We use Greek letters to label index of the Lorentz tensors, while Latin letters starting in the alphabet tables $a$, $b$, $\cdots$ to represent conserved charges, intermediate letters $i$, $j$, $k$, $l$ to represent gluon and different flavors of quarks, $r$ and $m$ for the energy weighting in the momentum integrals.
\item
    Numbers: 
    $N$ is number of particle species, $N'$ is number of conserved charges, and $N_s$ is the order of $s$-rank tensors kept in the moment expansion.
\item
    Projected tensors: 
    We used mostly negative signature for the metric.
    We denote the spatial projection operator in the fluid rest frame as $\Delta^{\mu\nu} \equiv g^{\mu\nu} - u^\mu u^\nu$. We then define the following notations the projected tensors $X^{\langle\mu_1\cdots\mu_s\rangle} \equiv \Delta^{\mu_1\cdots\mu_s}_{\nu_1\cdots\nu_s} X^{\nu_1\cdots\nu_s}$, the symmetrized tensors $X^{(\mu\nu)} \equiv (X^{\mu\nu} + X^{\nu\mu})/2$ and the anti-symmetrized tensors $X^{[\mu\nu]} \equiv (X^{\mu\nu} - X^{\nu\mu})/2$.
\item 
    Derivatives:
    We introduce the co-moving time derivative $D=u^\mu\partial_\mu$ and spatial derivative $\nabla^\mu\equiv\Delta^{\mu\nu}\partial_\nu$. Then, one can define the expansion rate $\theta\equiv\nabla_\mu u^\mu$, the shear stress tensor $\sigma^{\mu\nu}\equiv\nabla^{\langle\mu}u^{\nu\rangle}$, and the vorticity tensor $\omega^{\mu\nu} = \nabla^{[\mu}u^{\nu]}$.
\item
    Momentum integrals:
    We use the following shorthand for the momentum integrals $\int_p [\cdot] \equiv \int \frac{d^3p}{(2\pi)^3p^0} [\cdot]$, and $\int_{p_1,\cdots,p_s} [\cdot] \equiv \int \Big(\prod_{i=1}^s \frac{d^3p_i}{(2\pi)^3p_i^0}\Big) [\cdot]$.
\item
    We also introduced the inverse temperature $\beta\equiv\frac{1}{T}$, scaled chemical potentials $\alpha_a\equiv\frac{\mu_a}{T}$, and the co-moving energy for a particle with momentum $p$, $E_p\equiv u\cdot p$.
\end{itemize}


\section{multicomponent Hydrodynamic Equations with Inelastic Reactions}\label{sec.hydro_equations}

\subsection{The Bridge Between Microscopic and Macroscopic Theories}
The hydrodynamic theory is a macroscopic theory to describe the dynamical evolution of a system based on the conservation of energy, momentum, and all conserved charges,
\begin{align}
\partial_\mu T^{\mu\nu}=0,\quad \partial_\mu N^\mu_a=0,\quad a=1,2,\cdots, N',
\label{eq.conservation}
\end{align}
as well as the second law of thermodynamics,
\begin{align}
    \partial_\mu S^\mu \geq 0\,.
    \label{eq.entropy}
\end{align}
$N'$ is the number of conserved charges, which is less than or equal to the number of components, $N'\leq N$.
Here, $T^{\mu\nu}$ and $N^\mu_a$ are respectively the energy-momentum tensor and charge currents. They are functions of space-time coordinate $(x)$ and can be decomposed into
\begin{align}
\begin{split}
N^\mu_a =\;& 
	n_a\,u^\mu + V^\mu_a , 
\end{split}\label{eq.N_hydro}
\\
\begin{split}
T^{\mu\nu} 
=\;&
	\varepsilon\, u^\mu u^\nu 
	- (P+\Pi)\, \Delta^{\mu\nu}
	+ \pi^{\mu\nu}
\\&
	+ W^\mu u^\nu + W^\nu u^\mu,
\end{split}\label{eq.T_hydro}
\end{align}
Here, $u^\mu$ is the four-velocity of the fluid, $\Delta^{\mu\nu} \equiv g^{\mu\nu} - u^\mu u^\nu$ is the spatial projection operator in the fluid rest frame, $n_a$ is the $a$th charge density, $\varepsilon$ the energy density, $P$ the pressure which determined via the equation of state, $P=P(\varepsilon,n_a)$. Meanwhile, $V^\mu_a$ is the diffusion current,  $\pi^{\mu\nu}$ the shear-viscous stress tensor,  $\Pi$ the bulk pressure and $W^\mu$ the heat flow. The latter four quantities are viscous corrections, characterizing how the system is away from local-equilibrium. They vanish when and only when Eq.~\eqref{eq.entropy} takes the equality, i.e., there is no entropy production. Additionally, the vector and tensor viscous components are defined to be orthogonal to velocity, i.e., $V^\mu_a u_\mu = W^\mu u_\mu = 0$, $\pi^{\mu\nu} u_\mu = 0$, and $\pi^{\mu\nu} u_\nu = 0$.

The time evolution of viscous terms is determined by the underlying microscopic process. In kinetic theory, the macroscopic hydrodynamic quantities can be expressed as the momentum integrals of the microscopic distribution function
\begin{align}
N^\mu_a(x) =\;& \sum_{k=1}^N d_k q_{ak}
	\int_p p^\mu f_k(x,p) \,,
\label{eq.N_kin}
\\
T^{\mu\nu}(x) 
=\;& \sum_{k=1}^N d_k
	\int_p p^\mu p^\nu f_k(x,p)\,,
\label{eq.T_kin}
\end{align}
where $k=1, 2, 3, \cdots, N$ labels the type of particle, $f_k = f_k(x,p)$ is the distribution function of the $k$th component, $d_k$ the degeneracy factor, $q_{ak}$ the $a$th charge carried by the $k$th component. 

To describe the evolution of the viscous terms, one would need to take into account the non-equilibrium corrections not only at the macroscopic level but also at the microscopic level.
Thus, we employ the method of moments developed in~\cite{Denicol:2012cn} and~\cite{DeGroot:1980dk} and express the distribution function as an equilibrium one plus corrections,
\begin{align}
f_k(x,p) =\;&
	f_k^{(0)}+ \delta f_k\,,\label{eq.moment_expansion}\\
\begin{split}
\delta f_k\equiv \;&
    f_k^{(0)}\tilde{f}_k^{(0)} \times\bigg[
	\sum_{n=0}^{N_0}  \mathcal{H}^{(0)}_{n}(E_p)\, \rho_{kn}(x)
\\&
	+\sum_{n=0}^{N_1} \mathcal{H}^{(1)}_{n}(E_p)\, \rho^{\mu}_{kn}(x) \,p_{\langle\mu\rangle}
\\&
	+\sum_{n=0}^{N_2} \mathcal{H}^{(2)}_{n}(E_p)\, \rho^{\mu\nu}_{kn}(x) \,p_{\langle\mu}\,p_{\nu\rangle}
	\bigg], 
\end{split}\label{eq.moment_expansion_1}
\end{align}
where $E_p\equiv u\cdot p$ is the co-moving energy for a particle with momentum $p$,  $f_k^{(0)} = \frac{1}{e^{ \beta\,u \cdot p -\alpha_k} + a}$ is the thermal distribution and $\tilde{f}_k^{(0)} = 1-a\,f_k^{(0)}$ is the statistical factor, with $a=1$($a=-1$) for fermions(bosons). Here, $\beta\equiv\frac{1}{T}$ and $\alpha\equiv\frac{\alpha}{T}$ are respectively the inverse temperature and scaled chemical potential. $\rho^{\mu_1 \cdots \mu_s}_{kn}$'s are the space-time dependent moments characterizing, locally, how the microscopic state is deviated from the thermal equilibrium.
$\mathcal{H}_{kn}^{(\ell)}$'s are polynomials of energy in the fluid co-moving frame. With their explicit forms and properties given in App.~\ref{app.moment_expansion}, one can find that
\begin{align}
    \rho^{\mu_1\cdots\mu_s}_{kn}\equiv
    d_k \int_p E^n_{p_k} p_k^{\langle\mu_1}\cdots p_k^{\mu_s\rangle} \delta f_k.
    \label{eq.moments}
\end{align}

It is worth noting that $f_k^{(0)}$ is the equilibrium correspondence of $f_k$, which can not be uniquely determined. A natural definition of $f_k^{(0)}$ is to ensure that it gives the same energy density and charge densities as $f_k$, i.e.,
\begin{align}
\begin{split}
\varepsilon(x)
\equiv \;&
	\sum_{k=1}^N d_k
	\int_p (u\cdot p)^2 f_k(x,p)
\\=\;&
	\sum_{k=1}^N d_k
	\int_p (u\cdot p)^2 f^{(0)}_k(x,p),
\end{split}\label{eq.matching_e}
\end{align}
and
\begin{align}
\begin{split}
n_a(x)
\equiv\;&
	\sum_{k=1}^N d_k q_{ak}
	\int_p (u\cdot p)\, f_k(x,p)
\\=\;&
	\sum_{k=1}^N d_k q_{ak}
	\int_p (u\cdot p)\, f^{(0)}_k(x,p) \,.
\end{split}\label{eq.matching_n}
\end{align}
With these matching conditions~\eqref{eq.matching_e} and~\eqref{eq.matching_n}, it is not hard to find that the viscous terms in the hydrodynamic equations can be expressed by the viscous moments that appeared in the microscopic distribution functions,
\begin{align}
\Pi =\;&
-\sum_{k}  \frac{m_k^2}{3}\rho_{k0},\label{eq.define_Pi}\\
V_a^{\mu} =\;&
\sum_{k} q_{ak}\rho^ {\mu}_{k0},\label{eq.define_V}\\
\pi^{\mu\nu} =\;&
\sum_{k} \rho^ {\mu\nu}_{k0},\label{eq.define_pi}
\end{align}

Now, the only remaining equilibrium quantity is the fluid velocity. The determination of $u^\mu$ is more subtle. Here, we adopt the common choice in high-energy physics that requires the velocity vector to be the eigenvector of the stress tensor,
\begin{align}
    T^{\mu\nu} u_\nu = \varepsilon\, u^\mu\,.
    \label{eq.matching_u}
\end{align}
This choice is refereed to as Landau frame, and it automatically leads to the vanishing of the heat flow, $W^\mu=0$. It is not hard to find that the matching conditions, \eqref{eq.matching_e}, \eqref{eq.matching_n}, and \eqref{eq.matching_u}, place the following constraints on the moments~\footnote{For completeness, we note that when taking the Eckart frame, the constraint~\eqref{matchu} would become $\sum_{k=1}^{N}q_{ak}\rho^\mu_{k0}=0$.},
\begin{align}
&\sum_{k=1}^{N}\rho_{k2}=0, \label{matche}\\
&\sum_{k=1}^Nq_{ak}\rho_{k1}=0,\label{matchn}\\
&\sum_{k=1}^{N}\rho^\mu_{k1}=0.\label{matchu}
\end{align}

\subsection{Evolution Equations}
With the above preparations, now we move on to derive the evolution equations for the hydrodynamic quantities.
First, we obtain
the equations of motion for $n_a$, $\varepsilon$ and $u^\mu$ by simplifying the conservation equations~\eqref{eq.conservation},
\begin{align}
\label{Dn1}
Dn_a=\;&
    -n_a\theta-\partial_\mu V_a^\mu,  \\
\label{De1}
D\varepsilon=\;&
    -(\varepsilon+p+\Pi)\theta+\pi^{\mu\nu}\sigma_{\mu\nu},  \\
\label{Du1}
Du_\mu =\;&
    \frac{\nabla_\mu p-\Pi Du_\mu+\nabla_\mu\Pi-\Delta_{\mu\beta}\partial_\alpha\pi^{\alpha\beta}}{\varepsilon+P},
\end{align}
and hence those of $\alpha_a$ and $\beta$,
\begin{align}
\left(\!\!\!\begin{array}{c}
D\alpha_1 \\
\vdots \\
D\alpha_{N'} \\
D\beta \\
\end{array}\!\!\!\right)
 =\;& 
\mathcal{T}\cdot
\left(\!\!\!\begin{array}{c}
Dn_1 \\
\vdots \\
Dn_{N'} \\
D\varepsilon \\
\end{array}\!\!\!\right)\,,
\end{align}
where
\begin{align}
\mathcal{T} \equiv \;& 
\left(\!\!\begin{array}{ccccc}
\frac{\partial n_1}{\partial\alpha_1}  & 0 & 0 &\frac{\partial n_1}{\partial\beta}\\
0 & \ddots& 0&\vdots\\
0& \cdots &  \frac{\partial n_{N'}}{\partial\alpha_{N'}}&\frac{\partial n_{N'}}{\partial\beta}\\
\frac{\partial \varepsilon}{\partial\alpha_1}   &\cdots & \frac{\partial \varepsilon}{\partial\alpha_{N'}}  &\frac{\partial \varepsilon}{\partial\beta}\\
\end{array}\!\!\right)^{-1}\,,
\end{align}
and the derivative is carried out with respect to one variable with others fixed. In above equations, we have introduced the co-moving time derivative $D=u^\mu\partial_\mu$ and spatial derivative $\nabla^\mu\equiv\Delta^{\mu\nu}\partial_\nu$, as well as the expansion rate $\theta\equiv\nabla_\mu u^\mu$ and the shear stress tensor $\sigma^{\mu\nu}\equiv\nabla^{\langle\mu}u^{\nu\rangle}$.

To derive the evolution equation of the dissipative currents, one would need the evolution of the microscopic states, which is governed by the relativistic Boltzmann equation:
\begin{align}
\label{Boltzmann}
&p_k^\mu\partial_\mu f_k(x,p_k)=-\sum_{l=1}^{N}\mathcal{R}_{kl}[f_k],
\end{align}
with the collision operator containing both the gain term and the loss term:
\begin{align}
\label{Ckl}
\begin{split}
\mathcal{R}_{kl} 
\equiv\;& 
    \frac{d_l}{2} \sum_{i,j=1}^{N} 
    \int_{p_i,p_j,p_l} (f_if_j\tilde{f}_k\tilde{f}_l-f_kf_l\tilde{f}_i\tilde{f}_j) 
\\&\qquad\qquad\times
    W_{kl\to ij}(p_i,p_j|p_k,p_l),
\end{split}
\end{align}
where $d_l$ represents the degeneracy of $l$ particle, and the $1/2$ factor is introduced to avoid double-counting when $i \neq j$, and to compromise the symmetric factor when $i=j=k=l$ (single-component elastic scattering). $W_{kl\to ij}(p_i,p_j|p_k,p_l)$ is the transition rate describing the scattering process of $kl\to ij$ and will be abbreviated as $W_{kl\to ij}$ hereafter. In this work, we focus on two-to-two scatterings only and ignore the particle number changing reactions. When there is no reaction threshold, the last step can be safely implemented, which is exactly the case we will discuss. However, we do include inelastic scattering that changes particle's identity. To be explicit, reactive collisions --- such as $q+\bar{q} \to g + g$ --- and elastic scattering between different components --- like $q+g \to q + g$ --- are included.

Then, we implement the moment expansion~\eqref{eq.moment_expansion} and rewrite the Boltzmann equation as follow,
\begin{align}
\begin{split}
D\delta f_k =\;& -Df^{(0)}_k-E_{p_k}^{-1}p_k^\nu\nabla_\nu f^{(0)}_k 
\\&
	- E_{p_k}^{-1}p_k^\nu\nabla_\nu \delta f^{(0)}_k+E_{p_k}^{-1}\sum_{l}^{N}\mathcal{R}_{kl}[f_k].
\end{split}\label{Ddelta}
\end{align}
After a tedious but straightforward calculation, we obtain the motion equations for the moments of different types.
The motion equations for scalar, vector, and tensor moments respectively read
\begin{align}
\begin{split}
& D\rho_{kr} - \mathcal{R}_{k,r-1}
    - \alpha^{(0)}_{kr}\theta
\\=\;& 
    \Big(J^k_{r0}\sum_{a}^{N'}q_{ak}\mathcal{T}_{a,N'+1}
    -J^k_{r+1,0}\mathcal{T}_{N'+1,N'+1}\Big)
\\&\quad\times
    \Big(\Pi\,\theta - \pi^{\mu\nu}\sigma_{\mu\nu} \Big)
\\&
    +\sum_{a,b}^{N'}\Big(J^k_{r0}q_{ak}\mathcal{T}_{ab}
    -J^k_{r+1,0}\mathcal{T}_{N'+1,b}\delta_{ab}\Big)\partial_\mu V^\mu_a
\\&
    -\nabla_\mu \rho^{\mu}_{k,r-1}
    +r\rho^{\mu}_{k,r-1}Du_\mu
    +(r-1)\rho^{\mu\nu}_{k,r-2}\sigma_{\mu\nu}
\\&
    -\frac{\theta}{3}\Big((r+2)\rho_{kr}-(r-1)m_k^2\rho_{k,r-2}\Big)\,,
\end{split}\label{eq.eom.scalar}
\end{align}
\begin{align}
\begin{split}
&   D\rho_{kr}^{\langle \mu \rangle }
    -\mathcal{R}_{k,r-1}^{\langle\mu\rangle}
    +\sum_{a}^{N'}\alpha^{(1)}_{a,kr}\nabla^\mu \alpha_a
\\=\;&
    \frac{\beta J^k_{r+2,1}}{\varepsilon+P}\big(\Pi Du^\mu - \nabla^\mu\Pi + \Delta^\mu_\nu\partial_\lambda\pi^{\nu\lambda}\big) 
\\&
    +\frac{1}{3}\Big(r\, m_k^2\rho_{k,r-1}-(r+3)\rho_{k,r+1}\Big)Du^{\mu}
\\& 
    -\frac{1}{3}\nabla^{\mu}\big( m_k^2\rho_{k,r-1}-\rho_{k,r+1}\big) -\Delta _{\nu }^{\mu }\nabla _{\lambda}\rho_{k,r-1}^{\nu\lambda}
\\&
    +\frac{1}{3}\Big((r-1)m_k^{2}\rho_{k,r-2}^{\mu}
    -(r+3)\rho_{kr}^{\mu}\Big) \theta
\\& 
    +\frac{1}{5}\Big( (2r-2) m_k^{2}\rho_{k,r-2}^{\nu }
    -(2r+3)\rho_{kr}^{\nu}\Big) \sigma_{\nu}^{\mu}
\\&
    +\rho_{kr}^{\nu} {\omega^\mu}_{\nu} 
    - r\rho_{k,r-1}^{\mu \nu}Du_{\nu} 
    + (r-1)\rho_{k,r-2}^{\mu \nu \lambda }\sigma_{\nu \lambda},
\end{split}\label{eq.eom.vector}
\end{align}
and
\begin{align}
\begin{split}
&   D\rho_{kr}^{\langle \mu \nu \rangle}
    -\mathcal{R}_{k,r-1}^{\langle \mu \nu \rangle}
    -\alpha^{(2)}_{kr}\sigma ^{\mu \nu }
\\=&
    \frac{2}{15}\Big( (r-1)m_k^{4}\rho_{k,r-2}  -(2r+3) m_k^{2}\rho_{kr}
\\&
    +(r+4)\rho_{k,r+2}\Big) \sigma^{\mu\nu}
    +\frac{2}{5}\nabla^{\langle \mu} \big(\rho_{k,r+1}^{\nu\rangle}-m_k^{2}\rho_{k,r-1}^{\nu
	\rangle}\big)
\\&
	+2\rho_{kr}^{\lambda \langle \mu  }\omega _{{\,}%
	\lambda }^{\nu \rangle}
    -\frac{2}{5}\Big((r+5)\rho_{k,r+1}^{\langle\mu}-rm_k^{2}\rho_{k,r-1}^{\langle\mu}\Big)Du^{\nu\rangle}
\\&
    -\frac{1}{3}\Big( (r+4)\rho_{r}^{\mu \nu }-(r-1)m_k^{2}\rho_{k,r-2}^{\mu \nu }\Big) \theta
\\&
    -\frac{2}{7}\Big((2r+5)\rho_{kr}^{\lambda\langle\mu}
    -2(r-1)m_k^{2}\rho_{k,r-2}^{\lambda\langle\mu}\Big) 
    \sigma_{\lambda}^{\nu\rangle}
\\&
    +r\rho^{\mu\nu\lambda}_{k,r-1}Du_\lambda-\Delta^{\mu\nu}_{\alpha\beta}\nabla_\lambda \rho^{\alpha\beta\lambda}_{k,r-1}+(r-1)\rho^{\mu\nu\alpha\beta}_{k,r-2}\sigma_{\alpha\beta}.
\end{split}\label{eq.eom.tensor}
\end{align}
where $\omega^{\mu\nu} \equiv \nabla^{[\mu}u^{\nu]}$ is the vorticity tensor.
The pre-factors of the thermodynamic force are given by
\begin{align}
\begin{split}
\alpha^{(0)}_{kr} \equiv\; & 
    (1-r)I^k_{r1}
    -I^k_{r0} 
    +\Big(J^k_{r0}\sum_{a}^{N'}q_{ak}\mathcal{T}_{a,N'+1}
\\&
   -J^k_{r+1,0}\mathcal{T}_{N'+1,N'+1}\Big)(\varepsilon+P)
\\&
    +\sum_{a,b}^{N'}\Big(J^k_{r0}q_{ak}\mathcal{T}_{ab} -J^k_{r+1,0}\mathcal{T}_{N'+1,b}\delta_{ab}\Big)n_b\,,
\end{split}\label{eq.coef_alpha_0}\\
\alpha^{(1)}_{a,kr} \equiv\;& 
    q_{ak}J^k_{r+1,1}-\frac{n_aJ^k_{r+2,1}}{\varepsilon+P}\,,
\label{eq.coef_alpha_1}\\
\alpha^{(2)}_{kr} \equiv\;& 
   2 I^k_{r+2,1} + 2 (r-1)I^k_{r+2,2}\,.
\label{eq.coef_alpha_2}
\end{align}
whereas the thermodynamic functions are defined as
\begin{align}
\begin{split}
    I^k_{nq} \equiv\;& \frac{d_k 
    \int_{p} (E_p)^{n-2q} \Big(-\Delta_{\alpha\beta}p^{\alpha} p^{\beta}\Big)^{q}f^{(0)}_{k}}{(2q+1)!!}\,,
\end{split}  \label{eq.Inq} \\
\begin{split}
    J^k_{nq} \equiv\;&\frac{d_k \int_{p}
    (E_p)^{n-2q}\Big(-\Delta_{\alpha \beta }p^{\alpha }p^{\beta }\Big)^{q} f^{(0)}_{k}\tilde{f}^{(0)}_{k}}{(2q+1)!!} \,.\label{eq.Jnq}
\end{split}
\end{align}
In Eqs.~(\ref{eq.eom.scalar}--\ref{eq.eom.tensor}), we have defined the $s$-order collision kernel as
\begin{align}
\begin{split}
    \mathcal{R}^{\langle\mu_1\cdots\mu_s\rangle}_{k,r-1}
    \equiv\;&  
    d_k \int_{p_k}
	p^{\langle\mu_1}_k\cdots p^{\mu_s\rangle}_k E^{r-1}_{p_k}\sum_{l=1}^{N}\mathcal{R}_{kl}[f_k]\,.
\end{split}\label{ckr0}
\end{align} 
Substituting the moment expansion of the distribution function~\eqref{eq.moment_expansion_1} into Eq.~\eqref{Ckl}, and only keeping terms linear in the $\delta f$, (one can refer to e.g. \cite{Denicol:2012cn,Denicol2021} for detailed derivation), we obtain:
\begin{align}
\label{ckr1}
\begin{split}
\mathcal{R}^{\langle\mu_1\cdots\mu_s\rangle}_{k,r-1}
=\;&
    -\sum_{n=0}^{N_s}\sum_{l=0}^{N}\mathcal{A}^{kl(s)}_{rn} \rho^{\langle\mu_1\cdots\mu_s\rangle}_{ln}+
\mathcal{O}\big((\delta f)^2\big)
\end{split}
\end{align} 
where the collision integral matrix elements are defined as:
\begin{align}
\label{eq.collision_A}
\begin{split}
\mathcal{A}^{(s)kl}_{r,n}
\equiv\;&
    \mathcal{G}^{(s)kl}_{r,n}
    +\mathcal{B}^{(s)k}_{r,n}\delta_{kl}\,,
\end{split}\\
\label{eq.collision_B}
\begin{split}
\mathcal{B}^{(s)k}_{r,n}
\equiv\;&
	\sum_{i,j,l=1}^{N}\frac{d_k d_l}{4s+2}
	\int_{p_i,p_j,p_k,p_l}
	E^{r-1}_{p_k} \mathcal{H}^{(s)}_{n} (E_{p_k})
\\&\quad\times
	p^{\langle\mu_1}_k\cdots p^{\mu_s\rangle}_k 
	p_{k\langle\mu_1}\cdots p_{k\mu_s\rangle} 
\\&\quad\times
	W_{kl\to ij}f^{(0)}_k f^{(0)}_l \tilde{f}^{(0)}_i \tilde{f}^{(0)}_j  \,,
\end{split}\\
\label{eq.collision_G}
\begin{split}
\mathcal{G}^{(s)kl}_{r,n}
\equiv\;&	
	\frac{d_k}{4s+2}  \sum_{i,j=1}^{N}
	\int_{p_i,p_j,p_k,p_l}
	E^{r-1}_{p_k} \mathcal{H}^{(s)}_{n}(E_{p_l})
\\&\quad\times
	p^{\langle\mu_1}_k\cdots p^{\mu_s\rangle}_k
	p_{l\langle\mu_1}\cdots p_{l\mu_s\rangle} 
\\&\quad\times
	\Big[\;\,
	d_l W_{kl\to ij}f^{(0)}_k f^{(0)}_l \tilde{f}^{(0)}_i \tilde{f}^{(0)}_j 
\\&\qquad
	-d_i W_{ki \to lj}f^{(0)}_k f^{(0)}_i \tilde{f}^{(0)}_l \tilde{f}^{(0)}_j
\\&\qquad
	-d_j W_{kj \to il}f^{(0)}_kf^{(0)}_j\tilde{f}^{(0)}_i\tilde{f}^{(0)}_l\Big].
\end{split}
\end{align}
The exact form of the $\mathcal{A}^{(s)kl}_{r,n}$ elements dependents on the specific scattering cross-sections. With the derivations shown in App.~\ref{sec.A_matrix}, one can express them as the summation of repeated, low dimensional integrals. Therefore, a numerical calculation becomes plausible. Relevant results are given in Eqs.~(\ref{eq.B_result}--\ref{eq.G3_result}, \ref{eq.K_result}, \ref{eq.J5_result}, \ref{eq.J4_result}).

\subsection{Calculation of Transport Coefficients} \label{sec.transport_coefficients}

Eqs.~(\ref{eq.eom.scalar}--\ref{eq.eom.tensor}) is a equation set coupling different $\rho^{\mu_1\cdots\mu_s}$'s with each other. To learn the key features of the system evolution, such as what the Navier--Stokes limit is and how the moments approach such a limit, one needs to find the dynamical eigenmodes of the evolution equation set. This is plausible by focusing on leading terms in the Knudsen-number expansion and keeping up to $\mathcal{O}(\delta f)$ terms in the collisions kernel. In this subsection, we adopt the compact matrix form of the evolution equations and diagonalize the time evolution equations. Then, expressions for the transport coefficients and the relaxation times can be obtained.

We note that the particle index $k = 1, \cdots, N$, whereas the energy weighting $r = 0,\cdots,N_s$. For later convenience, we denote that $M_s \equiv (N_s+1)\,N$.
We can rewrite the dissipation equations in a compact matrix form, with $\rho_{kr}^{\langle\mu_1\cdots\mu_s\rangle}$ is the $(rN+k)$-th element of an $M_s$-dimensional vector $\boldsymbol{\rho}^{\langle\mu_1\cdots\mu_s\rangle}$, whereas $\mathcal{A}^{kl(s)}_{rn}$ the $(rN+k)$-th column and $(nN+l)$-th row of an $(M_s \times M_s)$-dimensional square matrix $\boldsymbol{\mathcal{A}}^{(s)}$.
The evolution equations~(\ref{eq.eom.scalar}--\ref{eq.eom.tensor}) can be re-expressed in a compact form
\begin{align}
\begin{split}
&	D \boldsymbol{\rho}^{\langle\mu_1\cdots\mu_s\rangle} 
	+ \boldsymbol{\mathcal{A}}^{(s)} \cdot \boldsymbol{\rho}^{\langle\mu_1\cdots\mu_s\rangle} 
\\=\;& 
	\sum_a \boldsymbol{\alpha}^{(s)}_a \mathrm{F}_a^{\langle\mu_1\cdots\mu_s\rangle} 
	+ \boldsymbol{H}^{\langle\mu_1\cdots\mu_s\rangle} \,,
\end{split}
\label{eq.matrix_evolution}
\end{align}
when $s=0$, $1$, and $2$, respectively. 
$\mathrm{F}$ is the thermodynamic force,
\begin{align}
\mathrm{F} \equiv\;&
	\theta \,,\\ 
\mathrm{F}_a^{\langle \mu \rangle} \equiv\;&
	\nabla^\mu \frac{\mu_a}{T}\,,\\
\mathrm{F}^{\langle\mu \nu\rangle}  \equiv\;&
	\sigma^{\mu\nu}\,,
\end{align}
The $a$-subscript is needed only for dissipative currents, whereas the charge summation $\sum_a$ is trivial when $s=0$ or $2$.
Finally, the $M_s$-dimensional vector $\boldsymbol{H}^{\langle\mu_1\cdots\mu_s\rangle}$ represents the higher-order terms in gradient expansion. Their explicit form can be obtained by comparing Eqs.~(\ref{eq.eom.scalar}--\ref{eq.eom.tensor}) with their compact form~\eqref{eq.matrix_evolution}. 
It shall be worth noting that the $\rho$'s in Eq.~\eqref{eq.matrix_evolution} are not independent of each other, they are constrained by the matching conditions~(\ref{matche}--\ref{matchu}). 
To implement the matching conditions, we introduce an $(M_s \times M_s)$-dimensional orthogonal matrix $\boldsymbol{P}_{(s)}$, so that 
\begin{align}
\boldsymbol{P}_{(s)} \cdot \boldsymbol{\rho}^{\langle\mu_1\cdots\mu_s\rangle} 
= \left(\begin{array}{c}
\rho_{\text{mat},1}^{\langle\mu_1\cdots\mu_s\rangle} 	\\
\vdots\\
\rho_{\text{mat},N^{(s)}_{\text{mat}}}^{\langle\mu_1\cdots\mu_s\rangle} 	\\
 \boldsymbol{\rho'}^{\langle\mu_1\cdots\mu_s\rangle} 
\end{array}
\right)\,,\label{eq.matching_matrix}
\end{align}
where $\rho_{\text{mat},i}^{\langle\mu_1\cdots\mu_s\rangle}$'s are the constraining conditions given in Eqs.~(\ref{matche}--\ref{matchu}), and $N^{(s)}_{\text{mat}}$ is the number of matching conditions.
There are $N^{(0)}_{\text{mat}}=N'+1$ conditions when $s=0$,
\begin{align}
\rho_{\text{mat},a} =\;& \sum_{k=1}^{N} q_{ak} \rho_{k1} \qquad a = 1,2,\cdots,N'\,,\\
\rho_{\text{mat},N'+1} =\;& \sum_{k=1}^{N} \rho_{k2} \,,
\end{align}
and one ($N^{(1)}_{\text{mat}}=1$) condition when $s=1$,
\footnote{
Equation~\eqref{eq.matching_1_landau} becomes
\begin{align}
\rho_{\text{mat},1}^\mu =\;& \sum_{k=1}^{N} q_{ak}\rho_{k0}^\mu\,,
\end{align}
when using the Eckart frame.}
\begin{align}
    \rho_{\text{mat},1}^\mu =\;& \sum_{k=1}^{N} \rho_{k1}^\mu\,.
    \label{eq.matching_1_landau}
\end{align}
 Meanwhile, there is no matching condition constraining the tensor components, i.e., $N^{(2)}_{\text{mat}}=0$.
We denote that
\begin{align}
M'_s \equiv M_s-N^{(s)}_{\text{mat}}\,,
\end{align}
and the relevant components $\boldsymbol{\rho'}^{\langle\mu_1\cdots\mu_s\rangle}$ forms a $M'_s$-dimensional vector. 

With details put in App.~\ref{sec.null_matching}, we show that components coupled with $\rho_{\text{mat},i}^{\langle\mu_1\cdots\mu_s\rangle}$ in the evolution matrix, the thermodynamic force, and higher-order terms vanish, which ensures the consistency between the evolution equation~\eqref{eq.matrix_evolution} and the matching conditions~(\ref{matche}--\ref{matchu}), 
\begin{align}
\boldsymbol{P}_{(s)} \cdot  \boldsymbol{\mathcal{A}}^{(s)}\cdot \boldsymbol{P}_{(s)}^T =\;& 
 	\left(\begin{array}{cccc}
		\boldsymbol{0}_{[N_\text{mat}^{(s)} \times N_\text{mat}^{(s)}]} &
		\boldsymbol{0}_{[N_\text{mat}^{(s)} \times M'_s]}  \\~\\
		\boldsymbol{b}_{[M'_s \times N_\text{mat}^{(s)}]} &
		  \boldsymbol{\mathcal{A}'}^{(s)}_{[M'_s \times M'_s]} \\
	 \end{array}\right) \,,
\label{eq.null_matching_A}\\
\boldsymbol{P}_{(s)} \cdot  \boldsymbol{\alpha}^{(s)} = \;&
 	\left(\begin{array}{c}
		\boldsymbol{0}_{[N_\text{mat}^{(s)} \times 1]}  \\~\\
		\boldsymbol{\alpha'}^{(s)}_{[M'_s \times 1]} \\
	 \end{array}\right) \,,
\label{eq.null_matching_alpha}\\
\boldsymbol{P}_{(s)} \cdot  \boldsymbol{H}^{\langle\mu_1\cdots\mu_s\rangle} = \;&
 	\left(\begin{array}{c}
		\boldsymbol{0}_{[N_\text{mat}^{(s)} \times 1]}  \\~\\
		\boldsymbol{H'}^{\langle\mu_1\cdots\mu_s\rangle}_{[M'_s \times 1]} \\
	 \end{array}\right) \,,
\label{eq.null_matching_H}
\end{align}
where the subscripts $[A\times B]$ indicate the dimensions of the sub-matrices.
We also note that the $\boldsymbol{b}$-components in matrix~\eqref{eq.null_matching_A} couple the dissipative terms to the matching conditions and thus do not contribute to their evolution. Hence, the relevant equation of motion is reduced as an $M'_s$-dimensional one,
\begin{align}
\begin{split}
&	D \boldsymbol{\rho'}^{\langle\mu_1\cdots\mu_s\rangle} 
	+ \boldsymbol{\mathcal{A}'}^{(s)} \cdot \boldsymbol{\rho'}^{\langle\mu_1\cdots\mu_s\rangle} 
\\=\;&
	 \sum_i \boldsymbol{\alpha'}^{(s)}_i \mathrm{F}_i^{\langle\mu_1\cdots\mu_s\rangle} 
	 + \boldsymbol{H'}^{\langle\mu_1\cdots\mu_s\rangle} \,.
\end{split}
\label{eq.matrix_evolution_reduced}
\end{align}
Obviously, the evolution of different elements $\boldsymbol{\rho'}^{\langle\mu_1\cdots\mu_s\rangle}$ couples with each other, and we obtain the dynamical eigenmodes by diagonalizing the evolution matrix $\boldsymbol{\mathcal{A}'}^{(s)}$,
\begin{align}
\begin{split}
&\boldsymbol{\Omega}_{(s)}^{-1} \cdot \boldsymbol{\mathcal{A}'}^{(s)} 
	\cdot \boldsymbol{\Omega}_{(s)} 
=
	\boldsymbol{\chi}^{(s)}
\\=\;& 
	\text{diag}\Big(\chi^{(s)}_0, \chi^{(s)}_1, \cdots, \chi^{(s)}_{M'_s-1} \Big)\,,
\end{split}\end{align}
where we have sorted the eigenvalues in ascending order: $0<\chi^{(s)}_0 \leq \chi^{(s)}_1 \leq \cdots \leq \chi^{(s)}_{M'_s-1}$.
We further employ the following notations for the dynamical eigenmodes,
\begin{align}
\boldsymbol{X}^{\langle\mu_1\cdots\mu_s\rangle}
\equiv \boldsymbol{\Omega}_{(s)}^{-1} \cdot \boldsymbol{\rho'}^{\langle\mu_1\cdots\mu_s\rangle}\,,
\label{eq.eigenmodes}
\end{align}
the pre-factors of the thermodynamic force terms,
\begin{align}
\boldsymbol{\beta}_i^{(s)} 
\equiv\;& 
	\boldsymbol{\Omega}_{(s)}^{-1}\cdot \boldsymbol{\alpha'}^{(s)}_i\,,
\end{align}
as well as the altered higher-order terms,
\begin{align}
\begin{split}
\boldsymbol{\widetilde{H}}^{\langle\mu_1\cdots\mu_s\rangle} 
\equiv \;&
	\boldsymbol{\Omega}_{(s)}^{-1} \cdot \boldsymbol{H'}^{\langle\mu_1\cdots\mu_s\rangle} 
\\&
	-\boldsymbol{\Omega}_{(s)}^{-1} \cdot  \Big(D \boldsymbol{\Omega}_{(s)} \Big) \cdot 
	\boldsymbol{X}^{\langle\mu_1\cdots\mu_s\rangle} \,.
\end{split}
\end{align}
Then, Eq.~\eqref{eq.matrix_evolution_reduced} is diagonalized as 
\begin{align}
\begin{split}
& D \boldsymbol{X}^{\langle\mu_1\cdots\mu_s\rangle} 
+ \boldsymbol{\chi}^{(s)} \cdot \boldsymbol{X}^{\langle\mu_1\cdots\mu_s\rangle} 
\\=\;&
 \sum_i \boldsymbol{\beta}_i^{(s)} \mathrm{F}_i^{\langle\mu_1\cdots\mu_s\rangle}
 + \boldsymbol{\widetilde{H}}^{\langle\mu_1\cdots\mu_s\rangle}
 \,.
\end{split}
\label{eq.matrix_evolution_diag}
\end{align}
So far, we have diagonalized the time evolution equations up to the first order in gradient expansion. Particularly, $\chi^{(s)}_i$ is the inverse of the relaxation time, characterizing how fast an eigenmode approaches its corresponding Navier--Stokes limit. 
We note that there are $M_0' = (N_0+1)N-N'-1$, $M_1' = (N_1+1)N-1$, and $M_2' = (N_2+1)N$ independent eigenmodes for zeroth, first, and second-order moments, respectively, whereas the corresponding number of dissipative quantities appeared in hydrodynamic equations~\eqref{eq.N_hydro} and~\eqref{eq.T_hydro} are respectively $1$, $N'$, and $1$. When and only when $N=N'=1$ and $N_s=2-s$, i.e., 14-moment formalism in single-component hydrodynamics, one would find equal numbers of dynamical eigenmodes and dissipative quantities. Otherwise, in multicomponent systems or single-component systems with keeping higher moments, there are more dynamical eigenmodes than the dissipative quantities of interest in hydrodynamics. In Ref.~\cite{Denicol:2012cn,Denicol2021} it is claimed that the redundant degrees of freedom correspond to rapidly decayed non-hydro modes, hence one can focus on the slowest modes that carry the charges of interest --- label by index $v$ --- and take the Navier--Stokes limit for the other modes, denoted by index $w$
\begin{align}
&X_{w}^{\langle\mu_1\cdots\mu_s\rangle} = 
	\sum_i \frac{\beta^{(s)}_{i,w}}{\chi^{(s)}_{w}} \mathrm{F}_i^{\langle\mu_1\cdots\mu_s\rangle}
	+ \frac{\widetilde{H}^{\langle\mu_1\cdots\mu_s\rangle}_w}{\chi^
	{(s)}_w}.
\end{align}
Taking such a hypothesis, the inversion of Eq.~\eqref{eq.eigenmodes} yields that
\begin{align}
\begin{split}
\rho_{u}'^{\langle\mu_1\cdots\mu_s\rangle} 
=\;& 
	\sum_{v} \Omega^{(s)}_{u,v} X_{v}^{\langle\mu_1\cdots\mu_s\rangle} 
\\&
	+ \sum_{w} \Omega^{(s)}_{u,w}
	\sum_i \frac{\beta^{(s)}_{i,w}}{\chi^{(s)}_{w}} \mathrm{F}_i^{\langle\mu_1\cdots\mu_s\rangle}
\\&
	+ \sum_{w} \Omega^{(s)}_{u,w} \frac{\widetilde{H}^{\langle\mu_1\cdots\mu_s\rangle}_w}{\chi^{(s)}_w}
	\,.
\end{split}
\end{align}

Denoting that 
\begin{align}
\boldsymbol{\Omega'}^{(s)}_{M_s \times M'_s}
\equiv {\boldsymbol{P}_{(s)}^T}_{[M_s \times M_s]}
\cdot \left(\begin{array}{c}
		\boldsymbol{0}_{[N_\text{mat}^{(s)} \times M'_s]}  \\~\\
		\boldsymbol{\Omega}^{(s)}_{[M'_s \times M'_s]}  \\
	 \end{array}\right)\,,
\end{align}
we can project back to the moments appearing in the distribution function,
\begin{align}
\begin{split}
\rho_{u}^{\langle\mu_1\cdots\mu_s\rangle} 
=\;& 
	\sum_{v} \Omega'^{(s)}_{u,v} X_{v}^{\langle\mu_1\cdots\mu_s\rangle} 
\\&
	+ \sum_{w} \Omega'^{(s)}_{u,w}
	\sum_i \frac{\beta^{(s)}_{i,w}}{\chi^{(s)}_{w}} \mathrm{F}_i^{\langle\mu_1\cdots\mu_s\rangle}
\\&
	+ \sum_{w} \Omega'^{(s)}_{u,w} \frac{\widetilde{H}^{\langle\mu_1\cdots\mu_s\rangle}_w}{\chi^{(s)}_w}
	\,.
\end{split} \label{eq.rho_slowest}
\end{align}
Here, we introduce the following shorthand for the pre-factors of the thermodynamic forces,
\begin{align}
\label{zeta}
\zeta \equiv \;& 
	\sum_{w=0}^{M'_0-1}  \sum_{k=1}^{N} 
	\frac{m_k^2}{3}\Omega'^{(0)}_{k0,w}\frac{\beta^{(0)}_{w}}{\chi^{(0)}_{w}} \,,
\\
\label{kappa}
\kappa_{ab}\equiv\;&
	\sum_{w=0}^{M'_1-1}  \sum_{k=1}^{N} q_{ak}
	\Omega'^{(1)}_{k0,w}\frac{\beta^{(1)}_{b,w}}{\chi^{(1)}_{w}} \,,
\\
\label{eta}
\eta \equiv \;&  
	\frac{1}{2} \sum_{w=0}^{M'_2-1}  \sum_{k=1}^{N} 
	\Omega'^{(2)}_{k0,w}\frac{\beta^{(2)}_{w}}{\chi^{(2)}_{w}} \,,
\end{align}
For scalar and tensor terms, only the slowest mode is relevant, whereas, for vector terms, the diffusion currents related to the conserved charges --- labeled by index $a$ --- in consideration are relevant. Substituting Eqs.~(\ref{eq.define_Pi}--\ref{eq.define_pi}) into Eq.~\eqref{eq.rho_slowest}, we find the relation between the dissipative quantities and their corresponding slowest eigenmode,
\begin{align}
\begin{split}
&	\Pi + \zeta\,\theta 
\\= \;&
	- \bigg(\sum_{k=1}^{N} \frac{m_k^2}{3} \Omega'^{(0)}_{k0,0}\bigg) 
		\,\bigg(X_{0} - \frac{\beta^{(0)}_0}{\chi^{(0)}_0} \theta-  \frac{\widetilde{H}_0}{\chi^{(0)}_0} \bigg) 
\\&
	- \sum_{k=1}^{N} \frac{m_k^2}{3} \sum_{w=0}^{M'_0-1}  \Omega'^{(0)}_{k0,w} \frac{\widetilde{H}_w}{\chi^{(0)}_w}\,,
\end{split}\label{eq.connect_slow_Pi}
\end{align}
\begin{align}
\begin{split}
&	V_a^{\mu} - \sum_b\kappa_{ab}\,\nabla^{\mu}\frac{\mu_b}{T} 
\\= \;&
	\sum_v \sum_{k=1}^{N} q_{ak}\Omega'^{(1)}_{k0,v}
	\bigg(X_{v}^{\langle\mu \rangle} - \sum_b\frac{\beta^{(1)}_{b,v}}{\chi^{(1)}_v} \nabla^{\mu}\frac{\mu_b}{T} - \frac{\widetilde{H}^{\langle \mu  \rangle}_v}{\chi^{(1)}_v} \bigg)
\\&
	+ \sum_{k=1}^{N} q_{ak} \sum_{w=0}^{M'_1-1}\Omega'^{(1)}_{k0,w} \frac{\widetilde{H}^{\langle \mu \rangle}_w}{\chi^{(1)}_w} \,,
\end{split}\label{eq.connect_slow_nu}
\end{align}
and
\begin{align}
\begin{split}
&	\pi^{\mu\nu} - 2\,\eta\,\sigma^{\mu\nu}  
\\= \;&
	\bigg(\sum_{k=1}^{N} \Omega'^{(2)}_{k0,0}\bigg)\; \bigg(X_{0}^{\langle\mu \nu \rangle}  - \frac{\beta^{(2)}_0}{\chi^{(2)}_0} \sigma^{\mu\nu} - \frac{\widetilde{H}^{\langle \mu \nu \rangle}_0}{\chi^{(2)}_0} \bigg)
\\&
	+ \sum_{k=1}^{N} \sum_{w=0}^{M'_2-1} \Omega'^{(2)}_{k0,w} \frac{\widetilde{H}^{\langle \mu \nu \rangle}_w}{\chi^{(2)}_w}
	\,,
\end{split}\label{eq.connect_slow_pi}
\end{align}

Before the very last step, we note that there are the moments in Eqs.~(\ref{eq.eom.scalar}--\ref{eq.eom.tensor}) with negative power index ($-r<0$). They can be linked to Eq.~\eqref{eq.rho_slowest} via
\begin{align}
\label{minusmode1}
\rho_{k,-r}^{\mu_1\cdots\mu_s}=\sum_{n=0}^{N_s}\mathcal{F}^{k(s)}_{r,n}\rho_{k,n}^{\mu_1\cdots\mu_s},
\end{align}
where  newly defined thermodynamic integral are introduced
\begin{align}
\mathcal{F}^{k(s)}_{r,n}
\equiv&\; \frac{s!\,d_k}{(2s+1)!(2\pi)^3} \int  \frac{\mathrm{d}^3 p_k}{p_k^0} f^{(0)}_k \tilde{f}^{(0)}_k
E^{-r}_{p_k}\nn\\
&\times\,\mathcal{H}^{(s)}_{p_k,n}(\Delta_{\alpha\beta}p^\alpha_kp^\beta_k)^s \label{Fkl}.
\end{align}

Before moving on to explicitly write down the second-order hydrodynamic equations, it is necessary to clarify the power counting and typical scales involved in the discussion. After that, the translation from the equations for moments to the hydrodynamic equations is unambiguous. The expansion scheme is closely connected with two expansion parameters: the Knudsen number $K_n$ and the inverse Reynolds number $R_F^{-1}$ where $F$ represents the dissipative quantities. The Knudsen number is defined as
\begin{align}
  K_n\equiv\frac{\ell_{\text{mfp}}}{L}  
\end{align}
measuring the relative strength of non-uniformity. Here $\ell_{\text{mfp}}$ is the mean free path and $L$ denotes the typical length associated with system non-uniformity. The Knudsen number expansion is equivalent to gradients expansion. The other parameter Reynolds number is used to quantify the deviation $\delta f$ from equilibrium $f_{\text{eq}}$, which are often estimated by the ratio of dissipative quantities to pressure or density, i.e.,
\begin{align}
 R^{-1}_{\Pi}=\frac{|\Pi|}{P},\quad
 R^{-1}_{V_a}=\frac{\sqrt{|V_a^\mu V_{a\mu}|}}{n_a},\quad
 R^{-1}_{\pi}=\frac{\sqrt{|\pi^{\mu\nu}\pi_{\mu\nu}|}}{P}.
\end{align}

Then, by substituting Eqs.~(\ref{eq.connect_slow_Pi}--\ref{eq.connect_slow_pi}) into Eq.~\eqref{eq.matrix_evolution_diag}, and keeping terms up-to second-order in Knudsen and inverse Reynolds numbers expansion, we can rewrite the motion equations of dissipative quantities as a relaxation equation 
\begin{align}
\tau_\Pi D{\Pi}+\Pi
=\;&
	-\zeta\,\theta+\mathcal{J}+\mathcal{K}+\mathcal{N},
\label{eq.EqofPi}\\
\sum_{b=1}^{N'}\tau_{ab} D{V}^{\langle\mu\rangle}_b+V^\mu_a
=\;&
	\sum_{b=1}^{N'}\kappa_{ab}\nabla^\mu\alpha_b+\mathcal{J}^\mu+\mathcal{K}^\mu+\mathcal{N}^\mu,\label{eq.EqofV}\\
\tau_\pi D{\pi}^{\langle\mu\nu\rangle}+\pi^{\mu\nu}
=\;&
	2\,\eta\,\sigma^{\mu\nu}+\mathcal{J}^{\mu\nu}+\mathcal{K}^{\mu\nu}+\mathcal{N}^{\mu\nu},
\label{eq.Eqofpi}
\end{align}
where the bulk and shear relaxation times are defined by their corresponding slowest mode,
\begin{align}
\tau_\Pi \equiv \frac{1}{\chi^{(0)}_0}\,,\qquad
\tau_\pi \equiv \frac{1}{\chi^{(2)}_0} \,,
\end{align}
where the relaxation for vector components involves $N'$ relevant eigenvalues and the relaxation time becomes a matrix,
\begin{align}
    \tau_{ab} \equiv \sum_{v=1}^{N'}  \frac{\widetilde{\Omega}_{av}(\widetilde{\Omega}^{-1})_{vb}}{\chi^{(1)}_v}\,,
\end{align}
where $\widetilde{\Omega}_{av} \equiv \sum_{k=1}^{N}q_{ak} \Omega'_{k0,v}$ is an $N'\times N'$ matrix and $\widetilde{\Omega}^{-1}$ is its inverse.
It is not hard to see the off-diagonal relaxation time, $\tau_{ab}$, is an odd function of $\mu_a$ and $\mu_b$. Therefore, in a neutral system, the relaxation time matrix would be diagonal and one can uniquely determine a relaxation time for each type of conserved charge.

The tensors $\mathcal{J}, \mathcal{J^\mu}$, and $\mathcal{J}^{\mu\nu}$ contain all terms of first-order in Knudsen and inverse Reynolds numbers,
\begin{align}
\mathcal{J}=&-\sum_a\Big(\,\ell_{\Pi V}\nabla\cdot V_a+\tau_{\Pi V}V^\mu_a\cdot\nabla_\mu P\\
+&\gamma_{\Pi V}V^\mu_a\cdot\nabla_\mu \beta+\sum_b\lambda_{\Pi V_{ab}}V^\mu_a\nabla_\mu\alpha_b\,\Big)
\nn\\-&\;
\delta_{\Pi\Pi}\Pi\theta+\lambda_{\Pi_\pi}\pi^{\mu\nu}\sigma_{\mu\nu},\\
\mathcal{J}^\mu=&-\tau_{V_a}V_{a,\nu}\omega^{\mu\nu}-\delta_{VV}V^\mu_a\theta-\ell_{V_a\Pi}\nabla^\mu\Pi
\nn\\+&\;
\ell_{V_a\pi}\Delta^{\mu\nu}\nabla_\alpha\pi^\alpha_\nu+\tau_{V_a\Pi}\Pi\nabla^\mu P-\tau_{V_a\pi}\pi^{\mu\nu}\nabla_\mu P
\nn\\-&\;
\lambda_{VV}V_{a\nu}\sigma^{\mu\nu}
+\sum_b\lambda_{V_{ab}\Pi}\Pi\nabla^\mu \alpha_b+\gamma_{V_a\Pi}\Pi\nabla^\mu \beta
\nn\\-&\;
\sum_b\lambda_{V_{ab}\pi}\pi^{\mu\nu}\nabla_\mu \alpha_b
-\gamma_{V_a\pi}\pi^{\mu\nu}\nabla_\nu \beta,\\
\mathcal{J}^{\mu\nu}=&\; 2\tau_\pi\pi^{\langle\mu}_\lambda\omega^{\nu\rangle\lambda} 
+\sum_b\bigg(\,\ell_{\pi V_b}\nabla^{\langle\mu} V^{\mu\rangle}_b
\nn\\-&\;
\tau_{\pi V_b}V^{\langle\mu}_b\nabla^{\nu\rangle} P+\gamma_{\pi V_b}V^{\langle\mu}_b\nabla^{\nu\rangle} \beta
\nn\\+&\;
\sum_{c}\lambda_{\pi V_{bc}} V^{\langle\mu}_b \nabla^{\nu\rangle}\alpha_c\,\bigg)-\delta_{\pi\pi}\pi^{\mu\nu}\theta
\nn\\+&\;
\lambda_{\pi_\Pi}\Pi\sigma^{\mu\nu}
-\tau_{\pi\pi}\pi^{\langle\mu}_\lambda\sigma^{\nu\rangle\lambda}.
\end{align}
The tensors $\mathcal{K}, \mathcal{K^\mu}$ and $\mathcal{K}^{\mu\nu}$ contain all terms of second-order in Knudsen number,
\begin{align}
\mathcal{K}=&\;\zeta_1\omega_{\mu\nu}\omega^{\mu\nu}+\zeta_2\sigma_{\mu\nu}\sigma^{\mu\nu}+\zeta_3\theta^2\nn\\
+&\sum_{b,c}\zeta_{4,bc}\nabla_\mu \alpha_b\nabla^\mu \alpha_c+\zeta_5\nabla_\mu P\nabla^\mu P\nn\\
+&\sum_b\zeta_{6,b}\nabla_\mu P\nabla^\mu \alpha_b+\zeta_7\nabla^2P+\sum_b\zeta_{8,b}\nabla^2 \alpha_b,\\
\mathcal{K}^\mu=&\;\kappa_1\sigma^{\mu\nu}\nabla_\nu P+\sum_b\kappa_{2,b}\sigma^{\mu\nu}\nabla_\nu \alpha_b\nn\\
+&\;\kappa_3\nabla^\mu P\theta+\sum_b\kappa_{4,b}\nabla^\mu \alpha_b\theta+\sum_b\kappa_{5,b}\omega^{\mu\nu}\nabla_\nu\alpha_b\nn\\
+&\;\kappa_6\Delta^\mu_\lambda\partial_\nu\sigma^{\lambda\nu}+\kappa_7\nabla^\mu\theta,\\
\mathcal{K}^{\mu\nu}=& \eta_1\omega^{\langle\mu}_\lambda\omega^{\nu\rangle\lambda} +\eta_2\theta\sigma^{\mu\nu}+\eta_3\sigma^{\langle\mu}_\lambda\sigma^{\nu\rangle\lambda}\nn\\
+&\;\eta_4\sigma^{\langle\mu}_\lambda\omega^{\nu\rangle\lambda}
+\sum_{b,c}\eta_{5,bc}\nabla^{\langle\mu}\alpha_b\nabla^{\nu\rangle}\alpha_c\nn\\
+&\;\eta_6\nabla^{\langle\mu}P\nabla^{\nu\rangle}P+\sum_b\eta_{7,b}\nabla^{\langle\mu}\alpha_b\nabla^{\nu\rangle}P\nn\\
+&\;\sum_b\eta_{8,b}\nabla^{\langle\mu}\nabla^{\nu\rangle}\alpha_b+\eta_9\nabla^{\langle\mu}\nabla^{\nu\rangle}P.
\end{align}
Finally, the tensors $\mathcal{N}, \mathcal{N}^\mu$ and $\mathcal{N}^{\mu\nu}$ contain all terms of second-order in inverse Reynolds number,
\begin{align}
&\mathcal{N}=\phi_1\Pi^2+\sum_{b,c} \phi_{2,bc}V^\mu_bV_{\mu,c}+\phi_3\pi^{\mu\nu}\pi_{\mu\nu},\\
&\mathcal{N}^\mu=\phi_{4}V_{a,\nu}\pi^{\mu\nu}+\phi_5\Pi V^{\mu}_a,\\
&\mathcal{N}^{\mu\nu}=\phi_6\Pi\sigma^{\mu\nu}+\phi_7\pi^{\langle\mu}_\lambda\pi^{\nu\rangle\lambda}+\sum_{b,c}\phi_{8,bc}V^{\langle\mu}_bV^{\nu\rangle}_c.
\end{align}
The explicit expression of second-order coefficients can be obtained by comparing the expression of Eq.~(\ref{eq.EqofPi}--\ref{eq.Eqofpi}) to the matrix expansion form. Hereafter, we mainly focus on the first-order coefficients.

\section{Transport Properties in Specific Systems}\label{sec.tran_prop}
With the framework developed in the precedent section, now we move on to compute the transport coefficients and relaxation times in specific systems. Noting that the collision integrals, namely Eqs.~(\ref{eq.B_result}--\ref{eq.G3_result}, \ref{eq.K_result}, \ref{eq.J5_result}, \ref{eq.J4_result}), can be simplified only for massless particles with Boltzmann distribution, we focus such a ideal scenario. In such a case, the bulk viscosity vanishes, $\Pi=0$.
In Sec.~\ref{sec.tran_prop.hard_sphere}, we will start from taking the hard-sphere cross-section. We vary the number of components, turn-on or off the elastic and/or inelastic scattering between them, to study the effect of different processes. Particularly, we will compare the multicomponent fluid with the degenerated single component fluid.
Finally, we will present the result, which implements the pQCD cross-sections, in Sec.~\ref{sec.tran_prop.pQCD}.

\subsection{Hard-Sphere Cross-sections}\label{sec.tran_prop.hard_sphere}
For the sake of simplicity, we start from the hard-sphere cross-section that $\sigma_{kl\to ij}(P)=const$ is independent of the momentum of the participating particles. For such a simplified case, the collision integrals can be computed exactly, see App.~\ref{app.hard_sphere}. We sequentially increase the complexity of the microscopic processes and study the effect of different components. In this subsection, we focus on the case that all chemical potentials vanish so that the conductivity and relaxation time matrices are diagonal.

\begin{figure}[!hbtp]
	\includegraphics[width=0.45\textwidth]{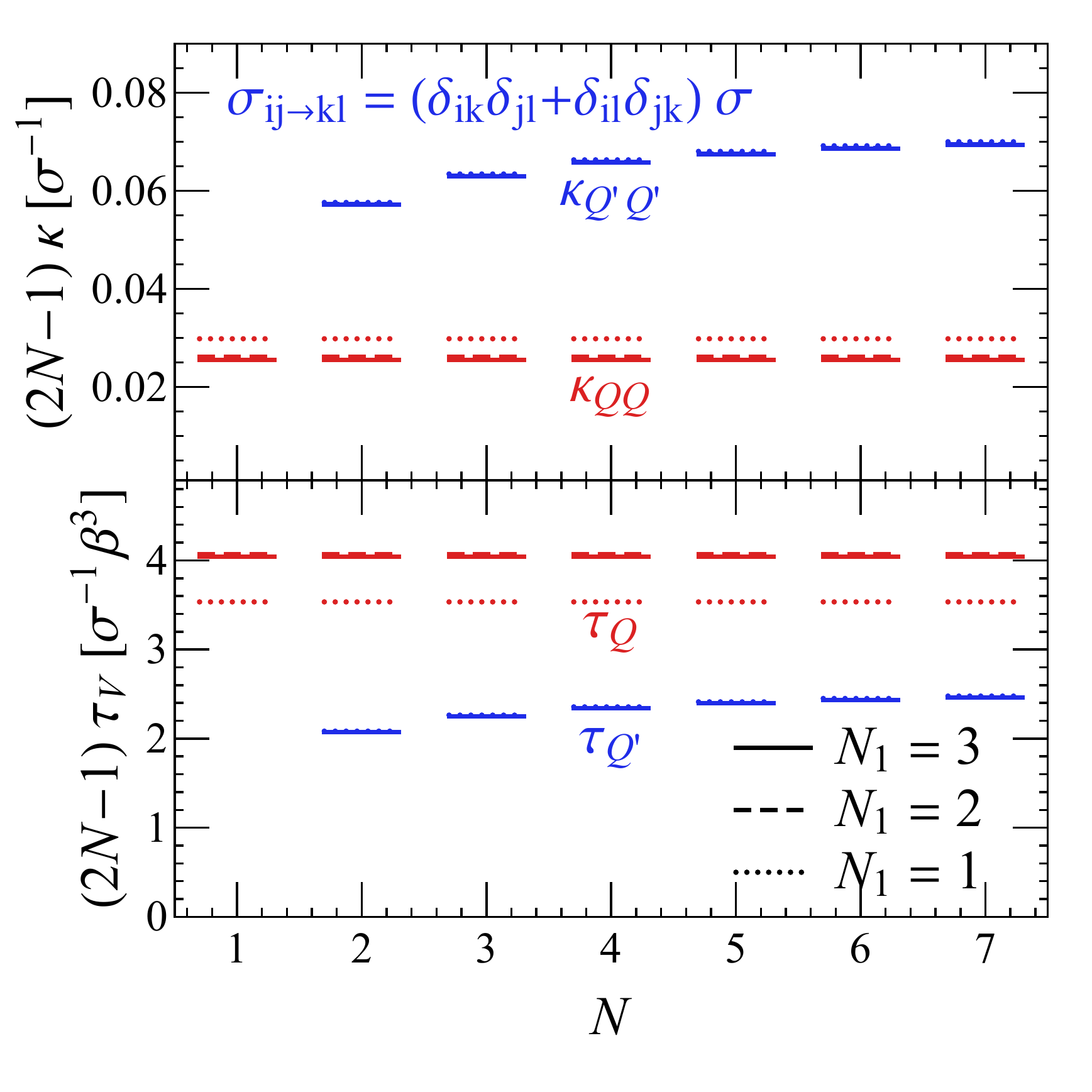} 
	\caption{Conductivity (upper panel) and relaxation times (lower panel) as a function of number of particle species.
	Dotted(dashed) lines are for $N_1=1$($N_1=2$). Red lines are for total charge $Q$, whereas blue lines correspond to the net charges $Q'$. }
	\label{fig.hardsphere_V_b}
\end{figure}
We start from an $N$-component system that only allows elastic scattering, and we assume identical cross-section between the same component, i.e., 
\begin{align}
    \sigma_{ij\to kl} = \delta_{ij}\delta_{jk}\delta_{kl}\, \sigma\,.
    \label{eq.hs_xsection_a}
\end{align}
In such a system, one can separate it as $N$ decoupled subsystems. Therefore, vector relaxation time ($\tau_V$), shear relaxation time ($\tau_\pi$), conductivity ($\kappa$), and shear-viscosity-to-entropy ratio ($\eta/s_{\mathrm{en}}$) remain independent of $N$. These value are listed in Table.~\ref{tab.hs_coefficients}. They are consistent with the corresponding single-component results in~\cite{Denicol:2012yr} (note that in this article, $\sigma$ stands for the differential cross-section whereas the same symbol is used to represent the total one. They are different by a factor of $2\pi$). 
\begin{table}[!hbt]
\begin{tabular}{c|c|c|c|c}
\hline\hline
$N_s+s-2$ & $\tau_V\,[\sigma^{-1}\beta^3]$ & $\tau_\pi\,[\sigma^{-1}\beta^3]$ & $\kappa\,[\sigma^{-1}]$ & $\eta/s_{\mathrm{en}}\,[\sigma^{-1}\beta^{2}]$ \\
\hline
0 &  $\frac{9\pi}{8}$   &  $\frac{5\pi}{6}$ &
    $\frac{3}{32\pi}$ & $\frac{\pi}{6}$\\
1 &  $1.295\pi$   &  $\pi$ &
    $\frac{21}{256\pi}$ & $\frac{7\pi}{44}$\\
2 &  $1.288\pi$  &  $\pi$ &  
   $\frac{0.08027}{\pi}$ & $\frac{163\pi}{1028}$\\
\hline\hline
\end{tabular}
\caption{Transport coefficients and relaxation times with hard-sphere interaction within the same particles~\protect{\eqref{eq.hs_xsection_a}}. \label{tab.hs_coefficients}}
\end{table}

Then we move on to an $N$-component system that only allows elastic scattering, and we assume identical cross-section between them, i.e.,
\begin{align}
    \sigma_{ij\to kl} = (\delta_{ik}\delta_{jl} + \delta_{il}\delta_{jk})\, \sigma\,.
    \label{eq.hs_xsection_b}
\end{align}
In such a system, every species of particle conserve on its own. Therefore, there are $N'=N$ conserve charges. We define the total charge as $Q\equiv N^{-1/2} \sum_k n_k$. There is a systematic way to generate the rest of the conserve charges. It is not hard to find that $Q$, $f_{\mathrm{norm}}\sum_k \cos \frac{2\pi\,a\,k}{N}$, and $f_{\mathrm{norm}}\sum_k \sin \frac{2\pi\,a\,k}{N}$, with $a$ being an integer and $f_{\mathrm{norm}}$ being the normalization factor, span a complete and orthogonal basis in the $N$-dimensional vector space. We find that the rest of them share the same values for relaxation time and conductivity, and their off-diagonal elements in the conductivity matrix always vanish. Therefore, we label them, uniformly, as $Q'$ when representing their transport properties. Noting that $Q'$'s measure the difference between different particles, they correspond to the conserved net charges such as baryon number, electric charge, or isospin.
We find that
$(2N-1) \tau_\pi$, $(2N-1) \eta/s_{\mathrm{en}}$, $(2N-1) \tau_Q$, and $(2N-1) \kappa_Q$ remain constant and takes the same values as listed in Table~\ref{tab.hs_coefficients}, whereas the change of $\tau_{Q'}$ and
$\kappa_{Q'}$ does not follow a simple formula. The latter are shown in Fig.~\ref{fig.hardsphere_V_b}. 

Third, we mimic the quantum statistics for indistinguishable particles, which let the cross-sections for elastic scattering between different particles reduce to half of that of the same particles, i.e.,
\begin{align}
    \sigma_{ij\to kl} = (\delta_{ij}\delta_{jk}\delta_{kl} +
    \delta_{ik}\delta_{jl} + \delta_{il}\delta_{jk})\, \frac{\sigma}{2}\,.
    \label{eq.hs_xsection_c}
\end{align}
We find that
$N\,\tau_\pi$, $N\,\eta/s_{\mathrm{en}}$, $N\,\tau_Q$, $N\,\kappa_Q$, $N\,\tau_{Q'}$, and $N\,\kappa_{Q'}$ remain independent of $N$. The value of the first four quantities are the same as those listed in Table~\ref{tab.hs_coefficients}, whereas the last two are listed in Table~\ref{tab.hs_coefficients_c}. It shall be worth emphasizing that one can treat a particle with $N$-degrees of freedom, such as spin and color, as a single-component system with degeneracy of $N$, correspondingly the effective cross-section shall be $N$ times of that of each channel. Change in the effective cross-section is also consistent with the fact that cross-section shall be summed over all final states. This also explains the $2N-1$ factor for the cross-section~\eqref{eq.hs_xsection_b}, which is the number of final state, i.e., one for the same particle scattering and two for different particles.

\begin{table}[!hbt]
\begin{tabular}{c|c|c}
\hline\hline
$N_s+s-2$ & $N\,\tau_{Q'}\,[\sigma^{-1}\beta^3]$ & $N\,\kappa_{Q'Q'}\,[\sigma^{-1}]$ \\
\hline
0 &  $0.841\pi$   & $\frac{4}{17\pi}$ \\
1 &  $0.837\pi$   & $\frac{0.234}{\pi}$ \\
2 &  $0.836\pi$   & $\frac{0.233}{\pi}$ \\
\hline\hline
\end{tabular}
\caption{Relaxation time and conductivity for net charges $Q'$ using hard-sphere potential cross-section~\protect{\eqref{eq.hs_xsection_c}}. \label{tab.hs_coefficients_c}}
\end{table}

Finally, we turn on reactions. 
In the exaggerated scenario that 
\begin{align}
    \sigma_{ij\to kl} = \sigma,
    \label{eq.hs_xsection_d}
\end{align}
for arbitrary $i$, $j$, $k$, and $l$, $N^3 \tau_\pi$, $N^3 \eta/s_\mathrm{en}$, $N^3 \tau_Q$, $N^3 \kappa_{QQ}$, $N^3 \tau_{Q'}$, $N^3 \kappa_{Q'Q'}$ are independent of $N$.
Combining the analysis using cross-sections (\ref{eq.hs_xsection_a}--\ref{eq.hs_xsection_d}), one can see that the effect of increasing the number of interaction channels --- either elastic or inelastic --- influences the transport properties by increasing the possible number of final states in each binary scattering. This again supports that one can treat different intrinsic degrees of freedom of a particle can be treated as the degenerated single-component system, as long as the transition between different states is properly reflected by the interaction cross-sections.

Among these cases, one could find the ordering of the relaxation times that $\tau_{Q'} < \tau_\pi < \tau_{Q} = \tau_{V}^\mathrm{single}$. Keeping in remind that $Q'$'s correspond to the conserved net charges like baryon number or isospin, such a multicomponent analysis is essential for a better estimation of the relaxation times of the moments of interest.

\subsection{Leading Order pQCD Cross-Sections}\label{sec.tran_prop.pQCD}
\begin{figure}[!hbtp]
    \includegraphics[width=0.4\textwidth]{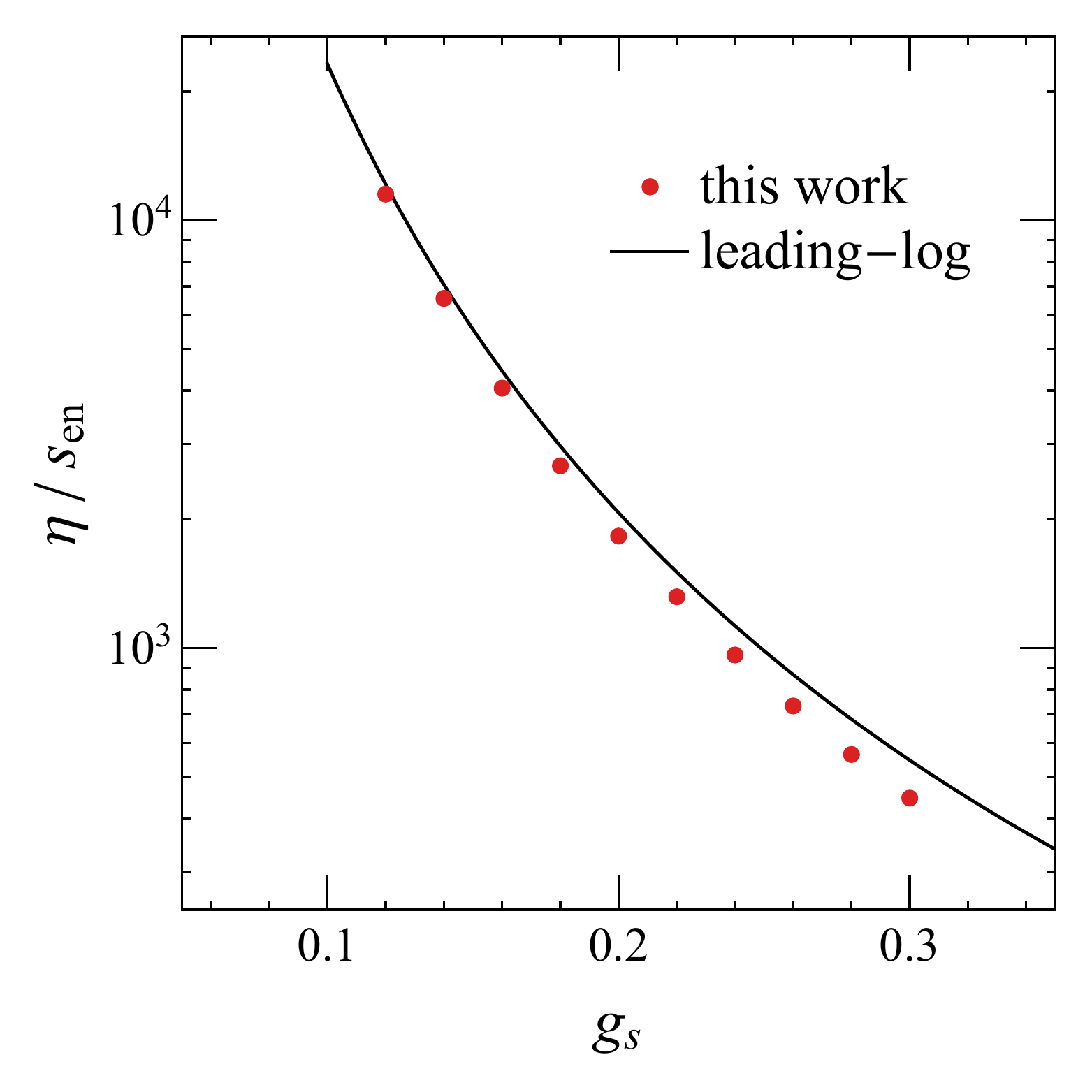}
	\caption{Entropy scaled shear viscosity as a function of strong coupling for a neutral system $\mu_B=\mu_I=0$. Red dots correspond to results in this work with $N_s+2-s=2$ and black curve represents the leading-log perturbative result~\protect{\cite{Arnold:2000dr}}.}
	\label{fig.leadinglog}
\end{figure}
\begin{figure}[!hbtp]
    \includegraphics[width=0.4\textwidth]{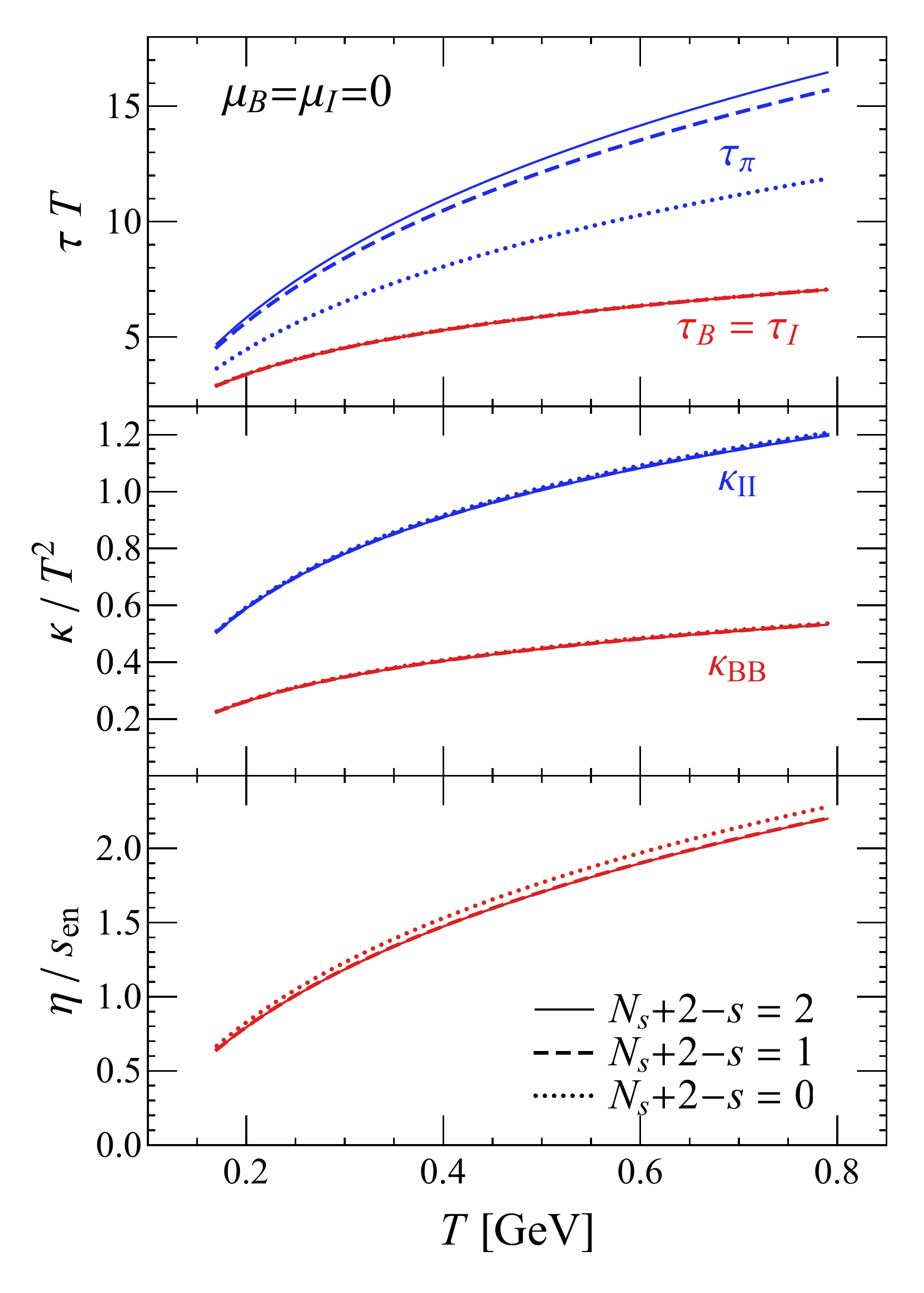}
	\caption{The relaxation times, conductivities and shear viscosity as a function of temperature for a neutral system $\mu_B=\mu_I=0$. Dotted, dashed, and solid lines respectively correspond to $N_s+2-s=0$, $1$, and $2$. In the upper panel, the relaxation times of net baryon diffusion and isospin diffusion degenerate, i.e., $\tau_B = \tau_I$.}
	\label{fig.zero_mu}
\end{figure}
\begin{figure*}[!htbp]
    \includegraphics[width=0.35\textwidth]{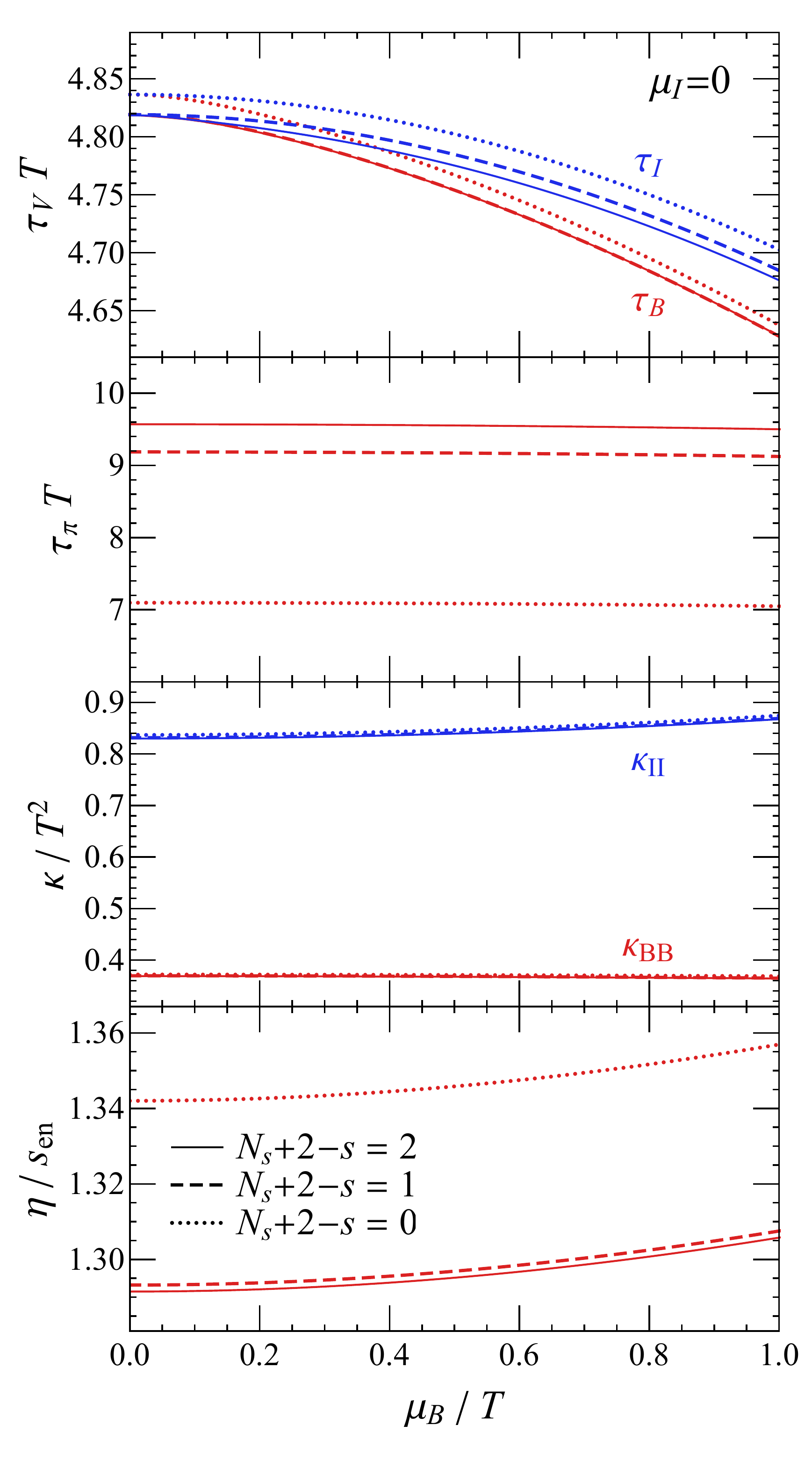}\qquad\qquad
    \includegraphics[width=0.35\textwidth]{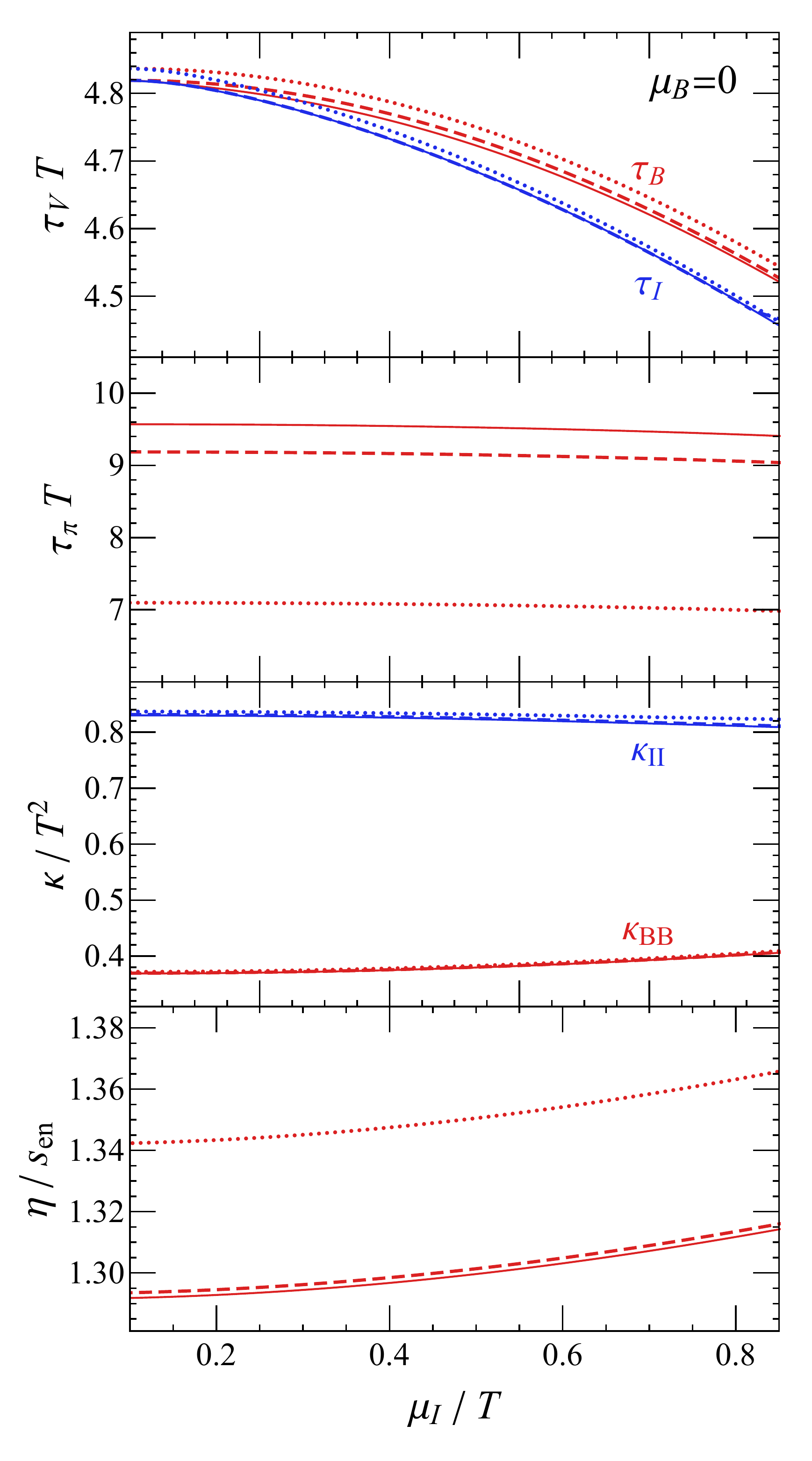}
	\caption{The relaxation times, conductivities, and scaled shear viscosity versus baryon(left) or isospin(right) chemical potential. Temperature is set to be $T=0.334$~GeV so that $\alpha_s(\mu_R^2=4\pi^2 T^2)=0.3$, whereas isospin(baryon) chemical potential vanishes in the left(right) panel. Dotted, dashed, and solid lines respectively correspond to $N_s+2-s=0$, $1$, and $2$. }
	\label{fig.zero_muIB}
\end{figure*}

\begin{figure}[!htb]
\includegraphics[width=0.4\textwidth]{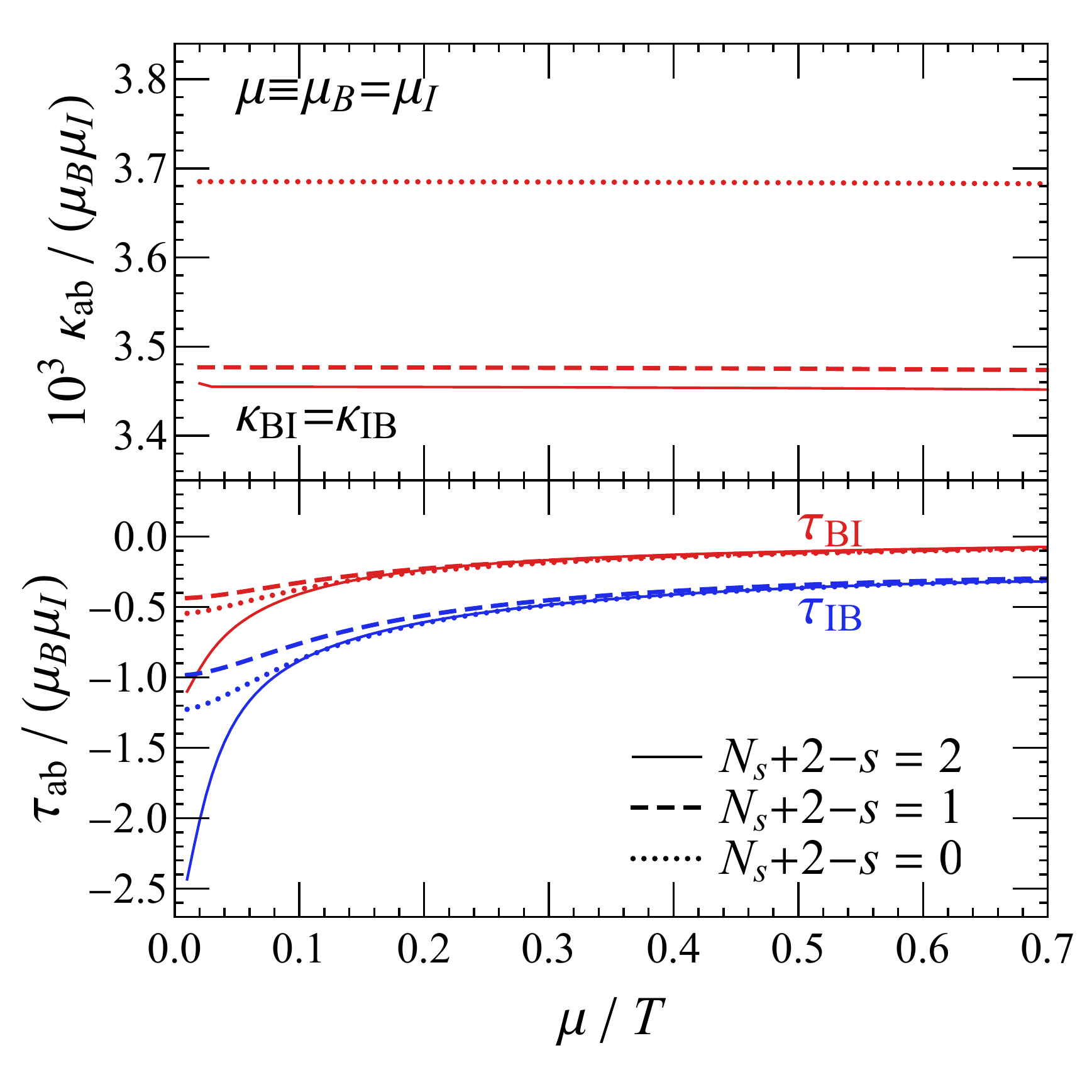} 
	\caption{The scaled off-diagonal diffusion coefficients as a function of the baryon or isospin chemical potential. Dotted, dashed, and solid lines respectively correspond to $N_s+2-s=0$, $1$, and $2$. }
	\label{fig.pQCD_nonzero_mu}
\end{figure}

We move on to compute the transport properties for a realistic system created in heavy-ion collisions. We consider a system consisting of gluon and $u$ and $d$ quarks and their antiquarks and implement the leading-order perturbative-QCD(LO pQCD) cross-section listed in Eqs.~(\ref{eq.pQCD_1}--\ref{eq.pQCD_11}).

With the LO pQCD cross-sections, the matrix elements $\mathcal{A}^{(s)kl}_{r,n}$  defined with Eqs.~(\ref{eq.B_result}--\ref{eq.G3_result}, \ref{eq.K_result}, \ref{eq.J5_result}, \ref{eq.J4_result}) can be evaluated following the procedures outlined in App.~\ref{sec.pQCD_cross}. 
In order to verify our calculation framework, let us take the $g_s \equiv \sqrt{4\pi\alpha_s} \ll 1$ limit and compare our transport coefficients with the analytical leading-log perterbative QCD result~\cite{Arnold:2000dr} that $\eta/s_\text{en} = \frac{86.5\,T^3}{g_s^4\ln g_s^{-1}} / \frac{40\times4T^3}{\pi^2}$.
Comparison is shown in Fig.~\ref{fig.leadinglog}, in which one can observe good agreement at the small $g_s$ limit. Our numerical calculation is more complete by keeping more than the leading-log terms, and therefore our result start to differ from~\cite{Arnold:2000dr} when $g_s$ increases.

Then we move on to the realistic temperature regime, in which the running coupling constant in the LO pQCD cross-sections is chosen to be  
\begin{equation}
    \alpha_s = \alpha_s(\mu^2_R=4\pi^2T^2),
    \label{eq.alpha_t}
\end{equation}
and evaluate the relaxation times and first-order transport coefficients.
The running of the coupling follows the beta function
$
\mu^2_R\frac{d\alpha_s(\mu^2_R)}{d\mu^2_R}=-(b_0\alpha_s^2+b_1\alpha_s^3+b_2\alpha_s^4),
$
with fixed point $\alpha_s(\mu_R^2 = 25 \,\text{GeV}^2)=0.2034$ and coefficients $b_0=\frac{33-2N_f}{12\pi}$,  $b_1=\frac{153-19N_f}{24\pi^2}$ and $b_2=\frac{2857-\frac{5033}{9}N_f+\frac{325}{27}N_f^2}{128\pi}$.

We first take the neutral limit ($\mu_B=\mu_I=0$) and compute the transport properties. With results exhibited in Fig.~\ref{fig.zero_mu}, several comments are in order:
\begin{itemize}
\item 
    First, we observe that the convergence of all these relaxation times and coefficients over the moment expansion with $N_s+2-s=0$, $1$, and $2$, which respectively correspond to $14$, $23$, and $32$ moment expansion for single-component fluid.
\item
    Second, all these temperature-scaled, dimensionless quantities increase with temperature since that the screening masses, $m_D$ and $m_q$, increase with temperature hence the effective couplings decrease.
\item
    Third, in the vanishing of chemical potentials, the transport properties of baryon and isospin degenerate, hence one finds $\tau_B=\tau_I$ and  $\kappa_{BB} = (4/9)\kappa_{II} $, with $4/9$ being the normalization factor. Meanwhile, we find that the off-diagonal elements of the conductivity matrix vanish in the neutral limit, $\kappa_{BI}=\kappa_{IB}=0$.
\item
    Fourth, similar to the hard-sphere potential case, it is observed that $\tau_V < \tau_\pi$. Particularly, we find the ratio between relaxation time and conductivity to be $\tau_B/\kappa_{BB} \approx 13\, \beta^3$, which is notably different from the single-component estimation $\tau_B/\kappa_{BB} \approx 158\, \beta^3$. This is meaningful in the phenomenological study and will be elaborated on in the Summary.
\item
    Last but not the least, the shear viscosity-to-entropy ratio is significantly higher than the holography limit~\cite{Kovtun:2004de}, $\frac{1}{4\pi}$, and the phenomenological results extracted from experimental observables invoking Bayesian inference~\cite{JETSCAPE:2020shq,JETSCAPE:2020mzn,Nijs:2020ors,Nijs:2020roc}. This is not surprising as LO pQCD cross-sections are based on a weakly coupling picture whereas the holography result takes the strongly coupling limit. Noting that the transport coefficients and the relaxation times are proportional to $\alpha_s^{-2}$, a naive extrapolation of the LO pQCD result to match the strongly coupling shear viscosity would result in the effective coupling to be $\alpha_s^\mathrm{eff} \sim 0.8$.
\end{itemize}

Then we move on to discuss the transport properties at non-vanishing chemical potentials. We fix the temperature as $T=0.334$~GeV, so that $\alpha_s(\mu_R^2=4\pi^2 T^2)=0.3$. In the left(right) panel of Fig.~\ref{fig.zero_muIB}, we let $\mu_I=0$($\mu_B=0$) and compute the relaxation times, conductivity, and shear viscosity as functions of $\mu_B$($\mu_I$). In addition to the convergence over moment expansion, we find that these transport properties are insensitive to the change of chemical potential.

We note that the off-diagonal elements of the conductivity and relaxation time matrices --- $\kappa_{BI}$, $\kappa_{IB}$, $\tau_{BI}$, and $\tau_{IB}$ --- are odd functions of both $\mu_B$ and $\mu_I$. They vanish when one of these chemical potentials vanishes. Therefore, we compute these off-diagonal elements at a non-vanishing chemical potential. For simplicity, we evaluate them at the line that $\mu_B = \mu_I$. As shown in Fig.~\ref{fig.pQCD_nonzero_mu}, we find $\kappa_{BI}=\kappa_{IB}$ and both of them are proportional to $\mu_B \mu_I$, and the proportionality constant is insensitive to chemical potential. The coincidence of $\kappa_{BI}$ and $\kappa_{IB}$ is known as the Onsager reciprocal relation, see e.g.~\cite{DeGroot:1980dk}. Meanwhile, $\tau_{IB}\approx (9/4)\tau_{BI}$. Regardless of their small value compared to the diagonal terms, the factors $\tau_{ab}/(\mu_B \mu_I)$ show higher sensitivity to the change in chemical potential.

\section{Summary and outlook}\label{sec.summary}
In this work, we derived the multicomponent second-order hydrodynamics equations with binary reactive processes, by exploiting the moment expansion formalism and truncating the Boltzmann collision kernel by keeping up to linear terms in non-equilibrium correction. 
We then derived the procedures to simplify the collision integrals of the evolution matrix with leading-order perturbative QCD cross-sections, which is essential for computing the relaxation times and first-order transport coefficients. For the pedagogical reason, we also reviewed the steps corresponding to hard-sphere potential.

Taking the hard-sphere interaction cross-section, we explored the effect of elastic scattering between different particles and inelastic binary reactions.
The inclusion of both these channels increases the frequency of interactions between two arbitrary particles, and therefore increases the effective cross-section and decreases the relaxation times, conductivity, and shear viscosity.

In a more realistic system taking LO pQCD cross-sections to describe the interactions between the components within QGP, we computed the dependence of relaxation times and transport coefficients on temperature and chemical potentials. As temperature increases, the Debye screening masses increase and, therefore, the effective cross-sections decrease. Thus, we observed that the relaxation times and transport coefficients increase with temperature. These coefficients exhibit weak dependence on the chemical potential, on the other hand.

Particularly, we found $\tau_B/\kappa_{BB} \approx 13\, \beta^3$ using the LO pQCD cross-section, which is notably different from the single-component estimation $\tau_B/\kappa_{BB} \approx 158\, \beta^3$. This shows how a multicomponent analysis is important for the realistic estimation of the relaxation times of moments of interest. In phenomenological hydrodynamics simulations, the relaxation time can hardly be constrained by experimental observables. One typically employs the relation between relaxation time and conductivity derived from relaxation time approximation(RTA) of Boltzmann equation~\cite{Denicol:2018wdp}, which takes the value that $\tau_B/\kappa_{BB} \approx \frac{3\alpha_B}{n_B} \approx 4 \, \beta^3$. The analysis in the current paper should provide a more sound and applicable guideline for phenomenological studies.

It shall be interesting to apply moment formalism to study how the collective hydrodynamic behavior develops in a far-from-equilibrium system. For such a purpose, two major improvements need to be made in the future. First, the current work focus on two-to-two scatterings, no matter elastic or inelastic. It shall be very interesting to consider number-changing processes, such as $gg \leftrightarrow ggg$, and elucidate their importance in momentum equilibration~\cite{Xu:2004mz}. Second, the current work truncates the collision kernel by keeping up to linear terms in the non-equilibrium expansion. It is important to keep the non-linear terms to study the far-from-equilibrium physics.

~\\\textbf{Note added.}  --- 
During the final preparation of the current manuscript, we noticed a recent, independent, but highly relevant study by J.~A.~Fotakis \textit{et al}~\cite{Fotakis:2022usk}, which also exploited the moment expansion method and derived the second-order multicomponent relativistic dissipative hydrodynamics. We arrived at the same results for the evolution equation for the general non-equilibrium moments [c.f. (\ref{eq.eom.scalar}--\ref{eq.eom.tensor}) in the current manuscript and Eqs.~(83--88) therein]. Then, Ref.~\cite{Fotakis:2022usk} explicitly derived the hydrodynamic equations for systems with baryon, electric charge, and strangeness being the conserved charges [Eqs.~(112--120) therein], whereas the current manuscript formally expresses the hydrodynamic equations for systems with an arbitrary number of components and conserved charges [Eqs.~(\ref{eq.EqofPi}--\ref{eq.Eqofpi})]. Results would be consistent if one specifies the underlying fluid constituent in the current work. In addition, the current work outlines the framework for computing the collision integrals with taking leading-order perturbative QCD cross-sections and further computes the relaxation times and transport coefficients. Related physics is also discussed.
\\\textbf{Acknowledgment} ---
The authors are grateful for helpful communications with Lipei Du and Jan A. Fotakis.
This work is  supported by the NSFC Grant Nos.11890710, 11890712, and 12035006 and the U.S. Department of Energy, Office of Science, Office of Nuclear Physics under grant No. DE-FG88ER40388.

\bibliographystyle{apsrev4-1}
\bibliography{paper.bib}

\clearpage
\begin{appendix}
\section{Moment Expansion}
\label{app.moment_expansion}

For pedagogical reason, we summarize the useful mathematical relations for moment expansion method, which is introduced in Ref.~\cite{DeGroot:1980dk,Denicol:2012cn,Denicol2021}. In such method, the distribution function $f$ is expanded around the local equilibrium distribution function $f_{0}$, 
\begin{equation}
f\equiv f_{0}+\delta f=f_{0
}\left( 1+\tilde{f}_{0}\,\phi\right) ,
\label{small_correc}
\end{equation}
where $\phi$ represents a general non-equilibrium correction.
$\phi$ is expanded in momentum
space with recourse to the irreducible tensors  composed of $1$, $p^{\langle \mu	\rangle}$, $p^{\langle \mu }p^{\nu \rangle} $, $p^{\langle \mu }p^{\nu}p^{\lambda \rangle}$
, $\cdots $, forming a complete and orthogonal tensor set with the minimum number of members, which are defined
by using the symmetrized traceless projections as 
\begin{equation}
    p^{\langle \mu_{1}} \cdots p^{\mu_{\ell}\rangle}
=\Delta^{\mu_{1} \cdots \mu_{\ell}}_{\nu_{1} \cdots \nu_{\ell}}
    p^{\nu_{1}}\cdots p^{\nu_{\ell}},
\end{equation}
where the projection tensor is given by
\begin{align}
\begin{split}
&   \Delta^{\mu_{1} \cdots \mu_{\ell}}_{\nu_{1} \cdots \nu_{\ell}}\\
\equiv\;&
    \sum_{n=0}^{[\ell/2]} (-4)^n \frac{n!\,(2\ell-2n)!}{(2\ell)!\,(\ell-n)!}
    \sum_{\{\mu,\nu\}}\\
& 
    \Big(\prod_{i=1}^{n} \Delta^{\mu_{2i-1},\mu_{2i}}\Delta_{\nu_{2i-1},\nu_{2i}} \Big)
    \Big(\prod_{i=2n+1}^{\ell} \Delta^{\mu_{i}}_{\nu_{i}} \Big),
\end{split}
\end{align}
where $[\ell/2]$ denotes the largest integer not bigger than $\ell/2$ and $\sum_{\{\mu,\nu\}}$ means summing all distinct permutations of $\mu$-type and $\nu$-type indices.

The tensors $p^{\langle \mu
	_{1}} \cdots p^{\mu_{\ell}\rangle}$ satisfy the
following orthogonality condition, 
\begin{align}
&\int_p\,d_{\text{gen}}\,
    F(E_p) p^{\langle \mu_{1}} \cdots p^{\mu_{\ell}\rangle} p_{\langle \nu_{1}} \cdots p_{\nu_{n}\rangle}  \nn \\
=&\frac{\ell!\,\delta_{\ell n}}{\left( 2\ell+1\right) !!}\Delta^{\mu_{1}\cdots \mu_{\ell}}_{\nu_{1} \cdots \nu_{\ell}}\int_p F(E_p)\left( \Delta^{\alpha
	\beta}p_{\alpha}p_{\beta}\right)^{\ell}.
	\label{orthogonality1}
\end{align}
Here $\ell$ and $n$ are non-negative integers, $d_{\text{gen}}$ is the degeneracy of particles, $F(E_p)$ is an arbitrary scalar function
of the co-moving energy $E_p\equiv u\cdot p$, and $\left( 2\ell+1\right) !!$ denotes the double
factorial.

Then the non-equilibrium
correction can be formally parametrized as \cite{Denicol:2012cn}, 
\begin{equation}
\phi=\sum_{\ell =0}^{\infty}\lambda^{\langle \mu_{1}\cdots \mu_{\ell}\rangle}(E_p)\,p_{\langle
	\mu_{1}}\cdots p_{\mu_{\ell}\rangle},
\label{expansion1}
\end{equation}
where the index $\ell $ represents the tensor rank. Furthermore, the coefficients $\lambda_{\mathbf{p}}^{\langle \mu_{1}\cdots \mu_{\ell}\rangle}$ can be expanded  by employing another orthogonal basis of functions $P_{n}^{(\ell)}(E_p)$, 
\begin{equation}
\lambda^{\langle \mu_{1}\cdots \mu_{\ell}\rangle
}(E_p)=\sum_{n=0}^{N_{\ell}}c_{n}^{\langle \mu_{1}\cdots \mu_{\ell
	}\rangle}P_{n}^{(\ell)}(E_p) \,, \label{expansion2}
\end{equation}
with $c_{n}^{\langle \mu_{1}\cdots \mu_{\ell}\rangle}$ to be determined later. Here $N_{\ell}$ denotes the number of
basis functions $P_{n}^{(\ell)}(E_p) $, 
\begin{equation}
    P_{n}^{(\ell)}(E_p)
=
    \sum_{r=0}^{n} a_{nr}^{(\ell)} E_{p}^{r},  \label{Poly}
\end{equation}
to meet the  orthonormality condition, 
\begin{equation}
\int_p
    \,d_{\text{gen}}\,\omega^{(\ell)}\,P_{m}^{(\ell)}(E_p) P_{n}^{(\ell)}(E_p)=\delta_{mn},
\label{conditions}
\end{equation}
 where the weight $\omega^{(\ell)}$ is defined as%
\begin{equation}
\omega^{(\ell)}= \frac{W^{(\ell)}}{%
	(2\ell +1) !!}\big( \Delta^{\alpha \beta}p_{\alpha}p_{\beta}\big)^{\ell} f_{0}\tilde{f}_{0}\;.
\end{equation}%
Here the coefficients $a_{nr}^{(\ell )}$ and the normalization constants $%
W^{( \ell )}$ can be found via Gram--Schmidt orthogonalization. First, by setting $P_0^{\ell}(E_p)=a^{\ell}_{00}=1$ without loss of generality, $W^{\ell}$ can be easily obtained from Eq.~\eqref{conditions}
\begin{align}
   W^{\ell}=(-1)^\ell \frac{1}{J_{2\ell,\ell}}.
\end{align}
Then we assume that $n\leq m$ and substitution of Eq.~\eqref{Poly} into Eq.~\eqref{conditions} leads to 
\begin{align}
    \sum^{m}_{n=0}\mathcal{D}^{(\ell m)}_{kn}\frac{a^{(\ell)}_{mn}}{a^{(\ell)}_{mm}}=\frac{J_{2\ell,\ell}}{(a^{(\ell)}_{mm})^2}\delta_{km},
\end{align}
where $\mathcal{D}^{(\ell m)}_{kn}\equiv J_{k+n+2\ell,\ell}$ and $k\leq m$.
Solve this equation and we will get
\begin{align}
    a^{(\ell)}_{mn} = J_{2\ell,\ell} (\mathcal{D}^{-1})^{(\ell m)}_{nm}.
\end{align}

Eventually, the single-particle distribution function can be expressed as 
\begin{equation}
    f
=
    f_{0} \Big(1 + \tilde{f}_{0} \sum_{\ell=0}^{\infty}\sum_{n=0}^{N_{\ell}} \mathcal{H}_{n}^{(\ell)}(E_p) \rho_{n}^{\mu_{1}\cdots \mu_{\ell}} p_{\langle \mu_{1}}\cdots p_{\mu_{\ell}\rangle}\Big) ,\ 
\label{main_f_expansion}
\end{equation}
with the function of energy-dependent
\begin{equation}
\mathcal{H}_{n}^{(\ell)}(E_p)
=
    \frac{W^{(\ell)}}{\ell!} \sum_{m=n}^{N_{\ell}} a_{mn}^{(\ell)} P_{m}^{(\ell)}(E_p)  \label{eq.Hk}
\end{equation}
and the generalized irreducible moment of $\delta f$, 
\begin{equation}
    \rho_{r}^{\mu_{1}\cdots \mu_{\ell}}
= 
    \int_p d_{\text{gen}}\,E_{p}^{r} \,p^{\langle \mu_{1}}\cdots p^{\mu_{\ell}\rangle} \delta f.  \label{rho}
\end{equation}

With all preparation works done, the expansion coefficients in Eq.~\eqref{expansion2}
can be immediately worked out by using Eqs.~(\ref{orthogonality1}) and (\ref{conditions}) and  given by 
\begin{align}
\begin{split}
c_{n}^{\langle \mu_{1}\cdots \mu_{\ell}\rangle}
\equiv\;&
    \frac{W^{(\ell)}}{\ell!}
    \int_p d_{\text{gen}}\,P_{
	n}^{(\ell)}(E_p)\,p^{\langle \mu_{1}}\cdots p^{\mu_{\ell}\rangle}\delta f
\\=\;& 
    \frac{W^{(\ell)}}{\ell!} \sum_{r=0}^{n} \rho_{r}^{\mu_{1}\cdots \mu_{\ell}} a_{nr}^{(\ell)}
\end{split}
\label{c_coeff}
\end{align}
for $n\leq N^{(\ell)} $.

\section{Prove of null-modes caused by matching conditions}
\label{sec.null_matching}

In this Appendix, we show that matching conditions correspond to trivial evolution equations, i.e., Eqs.~(\ref{eq.null_matching_A}--\ref{eq.null_matching_H}). In general, these relations can be straightforwardly obtained by applying moment expansion to the both side of Eq.~\eqref{Ddelta} and then take the transformation. Here we shown that they are satisfied when taking the explicit form, which is a cross-check of the results obtained. 

The matching conditions~\eqref{eq.matching_matrix} determine the following elements of the transformation matrix $\boldsymbol{P}_{s}$,
\begin{align}
P^{(0)}_{a,kr} =\;& \frac{q_{ak} \, \delta_{r,1}}{\sqrt{\sum_{j=1}^{N}(q_{aj})^2}}\,, \\
P^{(0)}_{N'+1,kr} =\;& N^{-\frac{1}{2}}  \,\delta_{r,2} \,,
\end{align}
and
\begin{align}
P^{(1)}_{1,kr} =\;& N^{-\frac{1}{2}}  \,\delta_{r,1} \,,
\end{align}
whereas the rest of elements are not fully determined --- the only constraints are they shall be orthogonal with each other and normalized to unity. Namely, they can be obtained by applying Gram--Schmidt orthogonalization.

We use tilde ( $\boldsymbol{\widetilde{\cdot}}$ ) to label vectors and tensors after the basis transformation,
\begin{align}
\boldsymbol{\widetilde{\mathcal{A}}}^{(s)} \equiv\;&
	\boldsymbol{P}_{(s)} \cdot  \boldsymbol{\mathcal{A}}^{(s)}
	\cdot \boldsymbol{P}_{(s)}^T\,,
\\
\boldsymbol{\widetilde{\alpha}}^{(s)} \equiv\;&
	\boldsymbol{P}_{(s)} \cdot  \boldsymbol{\alpha}^{(s)}\,,
\\
\boldsymbol{\widetilde{H}}^{\langle\mu_1\cdots\mu_s\rangle} \equiv\;&
	\boldsymbol{P}_{(s)} \cdot  \boldsymbol{H}^{\langle\mu_1\cdots\mu_s\rangle}\,.
\end{align}
Their elements can be written explicitly as
\begin{align}
\widetilde{\mathcal{A}}^{(s)}_{a,b} = \sum_{k,l}\sum_{r,n} P^{(s)}_{a,kr} P^{(s)}_{b,ln} \mathcal{A}^{kl(s)}_{rn}\,,
\end{align}
and
\begin{align}
\widetilde{\alpha}^{(s)}_{a} = \sum_{k}\sum_{r} P^{(s)}_{a,kr} \alpha^{(s)}_{kr}\,.
\end{align}

Implementing Eqs.~\eqref{eq.coef_alpha_0}, \eqref{eq.coef_alpha_1}, \eqref{eq.Inq}, and \eqref{eq.Jnq}, we find 
\begin{align}
\begin{split}
\alpha^{(0)}_{kr} \equiv\;& 
    (1-r)I^k_{r1} - I^k_{r0} + (\varepsilon+P)\times
\\&\quad
    \Big(J^k_{r0}\sum_{a}^{N'}q_{ak}\mathcal{T}_{a,N'+1}-J^k_{r+1,0}\mathcal{T}_{N'+1,N'+1}\Big)
\\&
+\sum_{a,b}^{N'}(J^k_{r0}q_{ak}\mathcal{T}_{ab}-J^k_{r+1,0}\mathcal{T}_{N'+1,b}\delta_{ab})n_b\,,
\end{split}
\end{align}
\begin{align}
\begin{split}
\widetilde{\alpha}^{(0)}_{a} 
= \;& 
    \Big(\sum_{j=1}^{N}(q_{aj})^2\Big)^{-\frac{1}{2}}
    \sum_k^N\bigg[ -q_{ak}I^k_{10} + (\varepsilon+P)\times
\\&\quad
	\Big(q_{ak}J^k_{10}\sum_{b}^{N'}q_{bk}\mathcal{T}_{a,N'+1}-q_{ak}J^k_{20}\mathcal{T}_{N'+1,N'+1}\Big)
\\&
	+q_{ak}\sum_{b,c}^{N'}(q_{bk}J^k_{10}\mathcal{T}_{bc} - J^k_{20}\mathcal{T}_{N'+1,c}\delta_{bc})n_c\,\bigg],
\end{split}\label{alphaa}
\end{align}
and
\begin{align}
\begin{split}
	\widetilde{\alpha}^{(0)}_{N'+1} 
= \;& \frac{1}{\sqrt{N}}\sum_k^N\bigg[
    -I^k_{21} -I^k_{20} + (\varepsilon+P)\times
\\&\quad
	\Big(J^k_{20}\sum_{a}^{N'}q_{ak}\mathcal{T}_{a,N'+1}-J^k_{30}\mathcal{T}_{N'+1,N'+1}\Big)
	\\&
	+\sum_{a,b}^{N'}(q_{ak}J^k_{20}\mathcal{T}_{ab}-J^k_{30}\mathcal{T}_{N'+1,b}\delta_{ab})n_b\,\bigg].
\end{split}\label{alphaN}
\end{align}

Proof of $\widetilde{\alpha}^{(0)}_{a}= \widetilde{\alpha}^{(0)}_{N'+1} =0$ would be cumbersome if one does not specify the number of components and conserved charges involved, which can be understood seeing the form of $\mathcal{T}$. For the sake of simplicity, we take two-flavor QGP as an example, which takes $N=5$, $N'=2$ and assumes that quarks and antiquarks are massive. Based on this setting, the Jaccobi transform matrix is 
\begin{align}
\begin{split}
\mathcal{T} =\;& 
\frac{1}{\frac{\partial n_1}{\partial \alpha_1}\frac{\partial n_2}{\partial \alpha_2}\frac{\partial  e}{\partial \beta}-\frac{\partial n_1}{\partial \alpha_1}\frac{\partial n_2}{\partial \beta}\frac{\partial  e}{\partial \alpha_2}-\frac{\partial n_1}{\partial \beta}\frac{\partial n_2}{\partial \alpha_2}\frac{\partial  e}{\partial \alpha_1}}\times
\\&
\left(\begin{array}{ccccc}
\frac{\partial n_2}{\partial \alpha_2}\frac{\partial  e}{\partial \beta}-\frac{\partial n_2}{\partial \beta}\frac{\partial  e}{\partial \alpha_2}  &\frac{\partial n_1}{\partial \beta}\frac{\partial  e}{\partial \alpha_2}  & -\frac{\partial n_1}{\partial \beta}\frac{\partial  n_2}{\partial \alpha_2}\\
\frac{\partial n_2}{\partial \beta}\frac{\partial  e}{\partial \alpha_1} & \frac{\partial n_1}{\partial \alpha_1}\frac{\partial  e}{\partial \beta}-\frac{\partial n_1}{\partial \beta}\frac{\partial  e}{\partial \alpha_1}& -\frac{\partial n_1}{\partial \alpha_1}\frac{\partial  n_2}{\partial \beta}\\
-\frac{\partial n_2}{\partial \alpha_2}\frac{\partial  e}{\partial \alpha_1} & -\frac{\partial n_1}{\partial \alpha_1}\frac{\partial  e}{\partial \alpha_2}& \frac{\partial n_1}{\partial \alpha_1}\frac{\partial  n_2}{\partial \alpha_2}\\
\end{array}\right) \,,
\end{split}
\end{align}
where indices $1$ and $2$ represent the net baryon number and isospin, respectively.

Recalling that thermodynamic functions $I^k_{nq}, J^k_{nq}$ is tightly linked with thermodynamic variables, e.g.,
\begin{align}
\quad&I^k_{1,0}=\frac{J^k_{2,1}}{T}=n_k,\quad I^k_{2,0}=\varepsilon_k,
\quad J^k_{3,1}=T(\varepsilon_k+P_k),
\end{align}
and some relations are useful:
\begin{align}
\quad& J^k_{n,q}=\frac{\partial I^k_{n,q}}{\partial \alpha_k},\\
\quad& J^k_{n,q}=-\frac{\partial I^k_{n-1,q}}{\partial \beta},\\
\quad& J^k_{n,q}=(n+1)T\,I^k_{n-1,q}.\label{Jnq}
\end{align}
Combining Eqs.(\ref{alphaa}--\ref{Jnq}) and performing the transfer from particle $k$ to charge $a$, a tedious but straightforward calculation leads to $\widetilde{\alpha}^{(0)}_{a}= \widetilde{\alpha}^{(0)}_{N'+1} =0$.

On the other hand,
\begin{align}
	\widetilde{\alpha}^{(1)}_{a,1}
=\;& N^{-\frac{1}{2}} \, \sum_{k=1}^{N} \Big(q_{ak}J^k_{2,1}-\frac{n_a J^k_{3,1}}{\varepsilon+P}\Big)\,,
\end{align}
We note that
\begin{align}
\varepsilon+P 	=\;& T^{-1}\sum_{k=1}^{N} J_{3,1}^k \,,\\
n_a 	=\;& T^{-1}\sum_{k=1}^{N} q_{ak} J_{2,1}^k\,.
\end{align}
Therefore, $\widetilde{\alpha}^{(1)}_{a,1} = 0$ is automatically fulfilled.

Meanwhile, the $\widetilde{\mathcal{A}}^{(s)}_{a,b}$ terms vanish because of matching conditions,
\begin{align}
&	\Bigg( \sum_{j=1}^{N}(q_{aj})^2\Bigg)^{\frac{1}{2}} \, 
	\sum_{k,r} P^{(0)}_{a,kr} \mathcal{A}^{(0)kl}_{rn}
\\=\;&
	\sum_{k=1}^{N} q_{ak} \mathcal{A}^{(0)kl}_{1n}
=
	q_{al} \mathcal{B}^{(0)l}_{1n} +
	\sum_{k=1}^{N} q_{ak} \mathcal{G}^{(0)kl}_{1n}=0,
\end{align}

\begin{align}
\begin{split}
&	N^{\frac{1}{2}} \sum_{k,r} P^{(0)}_{N'+1,kr} \mathcal{A}^{(0)kl}_{rn}
\\=\;&
	 \sum_{k=1}^{N} \mathcal{A}^{(0)kl}_{2n}
=
	\mathcal{B}^{(0)l}_{2n} +
	\sum_{k=1}^{N} \mathcal{G}^{(0)kl}_{2n}=0,
\end{split}
\end{align}

\begin{align}
\begin{split}
&	N^{\frac{1}{2}} \sum_{k,r} P^{(1)}_{1,kr} \mathcal{A}^{(1)kl}_{rn}
\\=\;&
	\sum_{k=1}^{N} \mathcal{A}^{(1)kl}_{1n}
=
	\mathcal{B}^{(1)l}_{1n} +
	\sum_{k=1}^{N} \mathcal{G}^{(1)kl}_{1n}=0.
\end{split}
\end{align}

With explicit form put in Eqs.(\ref{eq.B_result}--\ref{eq.G3_result}), we check that the above three conclusions hold.

\section{Calculation of the Collision Matrix Elements $\mathcal{A}^{(s)kl}_{r,n}$}\label{sec.A_matrix}

In this appendix, we simply the high-dimensional collision integrals, $\mathcal{A}^{(s)kl}_{r,n}$, into low-dimensional integrals that is realistic to be performed numerically. Noting its definition in Eq.~\eqref{eq.collision_A}, we separately discuss $\mathcal{B}^{(s)k}_{r,n}$ and $\mathcal{G}^{(s)kl}_{r,n}$ terms, which are defined in Eqs.~\eqref{eq.collision_B} and~\eqref{eq.collision_G}.

For carrying out the integration, we define the total and relative four-momentum:
\begin{align}
P^\mu\equiv\;&
    p^\mu_k+p^\mu_l=p^\mu_i+p^\mu_j\equiv P'^{\mu},\\
Q^\mu\equiv\;&
    \Delta^{\mu\nu}_P(p_{k\nu}-p_{l\nu}),\\
Q'^{\mu}\equiv\;&
    \Delta^{\mu\nu}_P(p_{i\nu}-p_{j\nu}),\\
\Delta^{\mu\nu}_P\equiv\;&
    \rho^ {\mu\nu}-\frac{P^\mu P^\nu}{P^2}.
\end{align}
With these definitions, the four momentum of participants can be expressed as:
\begin{align}
\begin{split}
&p^\mu_k=\frac{P^\mu + Q^\mu}{2},\qquad
p^\mu_l=\frac{P^\mu - Q^\mu}{2},\\
&p^\mu_i=\frac{P^\mu + Q'^\mu}{2},\qquad
p^\mu_j=\frac{P^\mu - Q'^\mu}{2},
\end{split}
\label{eq.PQ}
\end{align}
and these following relations will be useful:
\begin{align}
P\cdot Q = P\cdot Q'
=Q^2  + P^2
=Q'^2 + P^2=0.
\end{align}

\subsection{Splitting of the $\mathcal{B}^{(s)k}_{r,n}$ and  $\mathcal{G}^{(s)kl}_{r,n}$ integrals}
We note that $\mathcal{B}^{(s)k}_{r,n}$ and $\mathcal{G}^{(s)kl}_{r,n}$ are integrals of polynomials, and $\mathcal{B}^{(s)k}_{r,n}$ and $\mathcal{G}^{(s)kl}_{r,n}$ with different indices might share the same piece of monomial.
Therefore, we take the same notation as in Ref.~\cite{DeGroot:1980dk} and use the following short-hands for the mononial integrals,
\begin{align}
\begin{split}
\mathcal{J}^{(a,b,d,e,f)}_{kl \to ij}
\equiv\;&
	(2\pi)^{12}\int_{p_i,p_j,p_k,p_l}
	(P^2)^{a-1} e^{-\beta P\cdot u} 
\\&\times
	(u\cdot P)^b (u\cdot Q)^d(u\cdot Q')^e(Q\cdot Q')^f 
\\&\times
	W_{kl\to ij}(P,-\hat{Q}\cdot\hat{Q}'),
\end{split}\label{eq.intJ_B}\\
\begin{split}
\mathcal{K}^{(a,b,d,e,f)}_{kl \to ij}
\equiv\;&	
	(2\pi)^{12}\int_{p_i,p_j,p_k,p_l}
	(P^2)^{a-1} e^{-\beta P\cdot u} 
\\&\times
	\frac{(u\cdot P)^b(u\cdot Q)^d(u\cdot Q')^e(Q\cdot Q')^f}{u\cdot P+u\cdot Q}
\\&\times
	W_{kl\to ij}(P,-\hat{Q}\cdot\hat{Q}'),
\end{split}\label{eq.intK}\\
\begin{split}
\mathcal{J}^{(a,b,e,f)}_{kl \to ij}
\equiv\;&	
	(2\pi)^{12}\int_{p_i,p_j,p_k,p_l}
	(P^2)^{a-1} e^{-\beta P\cdot u} 
\\&\times
	\frac{(u\cdot P)^b(u\cdot Q')^e (Q\cdot Q')^f}{u\cdot P+u\cdot Q}
\\&\times
	W_{kl\to ij}(P,-\hat{Q}\cdot\hat{Q}').
\end{split}\label{eq.intJ_G}
\end{align}

We also note that the $\mathcal{G}$ integral~\eqref{eq.collision_G} is the sum of three terms that is of different characteristic, we split it into three terms
\begin{align}
\mathcal{G}^{(s)kl}_{r,n}
\equiv\;&
    \mathcal{G}1^{(s)kl}_{r,n}+\mathcal{G}2^{(s)kl}_{r,n}+\mathcal{G}3^{(s)kl}_{r,n}\,,
\\
\label{eq.collision_G1}
\begin{split}
\mathcal{G}1^{(s)kl}_{r,n}
\equiv\;&	
	\frac{d_k}{4s+2}  \sum_{i,j=1}^{N}
	\int_{p_i,p_j,p_k,p_l}
	E^{r-1}_{p_k} \mathcal{H}^{(s)}_{n} (E_{p_l})
\\&\quad\times
	p^{\langle\mu_1}_k\cdots p^{\mu_s\rangle}_k
	p_{l\langle\mu_1}\cdots p_{l\mu_s\rangle} 
\\&\quad\times
	d_l W_{kl\to ij}f^{(0)}_k f^{(0)}_l \tilde{f}^{(0)}_i \tilde{f}^{(0)}_j \,,
\end{split}\\
\label{eq.collision_G2}
\begin{split}
\mathcal{G}2^{(s)kl}_{r,n}
\equiv\;&	
	-\frac{d_k}{4s+2}  \sum_{i,j=1}^{N}
	\int_{p_i,p_j,p_k,p_l}
	E^{r-1}_{p_k} \mathcal{H}^{(s)}_{n}(E_{p_l})
\\&\quad\times
	p^{\langle\mu_1}_k\cdots p^{\mu_s\rangle}_k
	p_{l\langle\mu_1}\cdots p_{l\mu_s\rangle} 
\\&\quad\times
	d_i W_{ki \to lj}f^{(0)}_k f^{(0)}_i \tilde{f}^{(0)}_l \tilde{f}^{(0)}_j\,,
\end{split}\\
\label{eq.collision_G3}
\begin{split}
\mathcal{G}3^{(s)kl}_{r,n}
\equiv\;&	
	-\frac{d_k}{4s+2}  \sum_{i,j=1}^{N}
	\int_{p_i,p_j,p_k,p_l}
	E^{r-1}_{p_k} \mathcal{H}^{(s)}_{n}(E_{p_l})
\\&\quad\times
	p^{\langle\mu_1}_k\cdots p^{\mu_s\rangle}_k
	p_{l\langle\mu_1}\cdots p_{l\mu_s\rangle} 
\\&\quad\times
	d_j W_{kj \to il}f^{(0)}_kf^{(0)}_j\tilde{f}^{(0)}_i\tilde{f}^{(0)}_l\,.
\end{split}
\end{align}

Computing the integrals according to Dirac--Fermi distribution would be complicated, here we follow the same procedure of~\cite{DeGroot:1980dk} and the Boltzmann limit, so that $\tilde{f}=1$. Then we can simplify the integrals by taking the following procedures:
\begin{itemize}
\item 
    Perform the momentum variable substitution according to Eq.~\eqref{eq.PQ} for $\mathcal{B}$ and $\mathcal{G}1$, whereas $p_k=\frac{1}{2}(P^\mu+Q^{\mu})$, $p_i=\frac{1}{2}(P^\mu-Q^{\mu})$, $p_l=\frac{1}{2}(P^\mu+Q'^{\mu})$ for $\mathcal{G}2$,
    and $p_k=\frac{1}{2}(P^\mu+Q^{\mu})$, $p_j=\frac{1}{2}(P^\mu-Q^{\mu})$, $p_l=\frac{1}{2}(P^\mu-Q'^{\mu})$ for $\mathcal{G}3$.
\item
    Expand the thermal polynomials $\mathcal{H}^{(s)}_{n}$ according to \eqref{eq.Hk}.
\item
    Exploit the summation relation for massless particles,
\begin{align}
\begin{split}
&	
    p^{\langle\mu_1}_k\cdots p^{\mu_s\rangle}_k 
    p_{k,\langle\mu_1}\cdots p_{k,\mu_s\rangle} 
\\=\;&
    (-1)^s\, F_s \, E_{p_k}^{2s}\,,
\end{split}\\
\begin{split}
&	
    p^{\langle\mu_1}_k\cdots p^{\mu_s\rangle}_k
    p_{l,\langle\mu_1}\cdots p_{l,\mu_s\rangle} 
\\=\;&
    \sum_{a=0}^{s}\, F_{s,a}\, (p_k \cdot p_l)^{s-a} \, E_{p_k}^{a} \,  E_{p_l}^{a}\,,
\end{split}
\end{align}
where $F_s \equiv \frac{s!(2s+1)}{(2s+1)!!}$, and the relevant coefficients are
\begin{align}
\begin{split}
&	F_{0,0} = F_{1,0} = F_{2,0} = 1,\quad F_{1,1}= -1,\\
&	F_{2,1} = -2,\quad	F_{2,2} = 2/3.
\end{split}
\end{align}
\item
    Implement the shorhand defined in \eqref{eq.intK}.
\end{itemize}
With these steps, we find the integrals to be
\begin{align}
\begin{split}
\mathcal{B}^{(s)k}_{r,n}
=\;&
	\frac{(-1)^s F_s}{2s+1}\frac{W_k^{(s)}}{s!}
	\sum_{i,j,l=1}^{N}\frac{d_k d_l e^{\alpha_k+\alpha_l}}{(2\pi)^{12}}
	\sum_{m=n}^{N_s}\sum_{m'=0}^{m}
\\&\times
	\bigg( \frac{a^{(s)}_{mn} a^{(s)}_{mm'}}{2^{r+2s+m'}}
	\sum_{b=0}^{r+2s+m'} \frac{(r+2s+m')!}{b!(r+2s+m'-b)!}	 
\\&\times
	\mathcal{K}_{kl\to ij}^{(1,b,r+2s+m'-b,0,0)}\bigg)\,,
\end{split}\label{eq.B_result}
\end{align}
\begin{align}
\begin{split}
\mathcal{G}1^{(s)kl}_{r,n}=\;&
	\frac{d_k d_l\,W_{l}^{(s)}\,e^{\alpha_k+\alpha_l}}{(2\pi)^{12}(2s+1)s!}
	\sum_{i,j=1}^{N} \sum_{t=0}^{s} \sum_{m=n}^{N_s}\sum_{m'=0}^{m}
\\&\times
	\bigg( \frac{F_{s,t} a_{mn}^{(s)} a_{mm'}^{(s)}}{2^{r+s+t+m'}} 
	\sum_{b=0}^{r+t} \sum_{c=0}^{m'+t} 
\\&\quad
	\frac{(r+t)!}{b!(r+t-b)!} \frac{(-1)^c (m'+t)!}{c!(m'+t-c)!} 
\\&\times
	\mathcal{K}_{kl\to ij}^{(s-t+1,m'+t+b-c,r+t-b+c,0,0)} \bigg)\,,
\end{split}\label{eq.G1_result}
\end{align}
\begin{align}
\begin{split}
\mathcal{G}2^{(s)kl}_{r,n}
=\;&
	-\frac{d_k d_l\,W_{l}^{(s)}}{(2\pi)^{12}(2s+1) s!} \sum_{i,j=1}^{N} e^{\alpha_k+\alpha_i} 
	\sum_{t=0}^{s} \sum_{m=n}^{N_s}
\\& \quad
	\sum_{m'=0}^{m} \bigg( \frac{F_{s,t} a_{mn}^{(s)} a_{mm'}^{(s)}}{2^{r+2s+m'}}\sum_{b=0}^{r+t} \sum_{c=0}^{m'+t} \sum_{d=0}^{s-t}
\\&\times
	 \frac{(r+t)!}{b! (r+t-b)!}  \frac{(m'+t)!}{c! (m'+t-c)!} 
\\& \times
     \frac{(s-t)!}{d! (s-t-d)!} 
\\& \times
	 \mathcal{K}_{ki \to lj}^{(s-t-d+1, b+c, r+t-b, m'+t-c, d)} \bigg)\,,
\end{split}\label{eq.G2_result}
\end{align}
and
\begin{align}
\begin{split}
\mathcal{G}3^{(s)kl}_{r,n}
=\;&
	-\frac{d_k d_l\,W_{l}^{(s)}}{(2\pi)^{12}(2s+1) s!}
	\sum_{i,j=1}^{N} e^{\alpha_k+\alpha_j} 
	\sum_{t=0}^{s} \sum_{m=n}^{N_s}
\\& \times
	\sum_{m'=0}^{m} \frac{F_{s,t}  a_{mn}^{(s)} a_{mm'}^{(s)}}{2^{r+2s+m'}}
	\sum_{b=0}^{r+t} \sum_{c=0}^{m'+t} \sum_{d=0}^{s-t}
\\& \quad
	\bigg((-1)^{m'+t-c+d}
	 \frac{(r+t)!}{b! (r+t-b)!}
\\& \times
	\frac{(m'+t)!}{c! (m'+t-c)!}  \frac{(s-t)!}{d! (s-t-d)!} 
\\& \times
	 \mathcal{K}_{kj \to li}^{(s-t-d+1, b+c, r+t-b, m'+t-c, d)}\bigg)\,.
\end{split}\label{eq.G3_result}
\end{align}
Here, $a_{mn}^{(s)}$ and $W^{(s)}_{k}$ are respectively the coefficients and thermal weights defined in App.~\ref{app.moment_expansion}, For massless particles, they read
\begin{align}
&	a_{00}^{(s)} = 1\,, \\
&	a_{11}^{(s)} = \frac{\beta}{\sqrt{2s+2}} \,,\qquad
	a_{10}^{(s)} = -\sqrt{2s+2} \,,\\
&	a_{22}^{(s)} = \frac{\beta^2}{2\sqrt{(s+1)(2s+1)}} \,,\qquad
	a_{21}^{(s)} = -2\beta\sqrt{\frac{s+1}{2s+1}} \,,\nn\\
&
	a_{20}^{(s)} = \sqrt{(s+1)(2s+1)} \,,\\
&	a_{33}^{(s)} = \frac{\beta^3}{6\sqrt{4+6s+10s^2}} \,,\nn\\
&	a_{32}^{(s)} = -\beta^2\sqrt{\frac{2+31s-15s^2}{2+70s-32s^2}},\quad
	a_{31}^{(s)} = \beta\sqrt{\frac{36+8s+s^2}{4}} ,\nn\\
&   a_{30}^{(s)} = -\sqrt{4+6s+10s^2}, 
\end{align}
for $s=0,1,2$ and the thermal weights are
\begin{align}
W^{(s)}_{k} = \frac{(-1)^s}{J^k_{2s,s}} = \frac{(-1)^s}{(2s)!!} \frac{2\pi^2}{d_k} \beta^{2s+2} e^{-\alpha_k}\,.
\end{align}

\subsection{Computing the $\mathcal{J}$ and $\mathcal{K}$ integrals}
So far, we have expressed $\mathcal{B}^{(s)k}_{r,n}$ and $\mathcal{G}^{(s)kl}_{r,n}$  into the summation of $\mathcal{K}^{(a,b,d+1,e,f)}_{kl \to ij}$, and in this subsection, we move on to lower the integration order of such terms.
From the definition (\ref{eq.intJ_B}--\ref{eq.intJ_G}), one can see
\begin{align}
\mathcal{J}^{(a,b,d,e,f)}_{kl \to ij} =\;& \mathcal{K}^{(a,b+1,d,e,f)}_{kl \to ij} + \mathcal{K}^{(a,b,d+1,e,f)}_{kl \to ij} \,,\\
\mathcal{J}^{(a,b,e,f)}_{kl \to ij} =\;& \mathcal{K}^{(a,b,0,e,f)}_{kl \to ij}\,,
\end{align}
and it is not hard to derive the relation that
\begin{align}\begin{split}
\mathcal{K}^{(a,b,d,e,f)}_{kl \to ij} 
=\;&	(-1)^d \mathcal{J}^{(a,b+d,e,f)}_{kl \to ij} 
\\&
    + \sum_{n=0}^{d-1} (-1)^n \mathcal{J}^{(a,b+n,d-1-n,e,f)}_{kl \to ij}\,.
\end{split}\label{eq.K_result}\end{align}
Therefore, we only need to calculation $\mathcal{J}^{(a,b,d,e,f)}_{kl \to ij}$ and $\mathcal{J}^{(a,b,e,f)}_{kl \to ij}$, and then $\mathcal{K}^{(a,b,d,e,f)}_{kl \to ij}$ follows. The four- and five-indices $\mathcal{J}$-integrals are computed following the procedures outlined in Ref.~\cite{DeGroot:1980dk}. With the procedures outlined in the rest of this subsection, we will simplify them from sixteen-dimensional integrals to one-dimensional ones, and results will be given in Eq.~\eqref{eq.J5_result} and \eqref{eq.J4_result}.
It might be worth noting the unit of the integrals are
\begin{align}
[\mathcal{J}^{(a,b,d,e,f)}] =\;& [E]^{2a+b+d+e+2f+2},\\
[\mathcal{J}^{(a,b,c,d)}] =\;& [E]^{2a+b+c+2d+1}.
\end{align}

\subsubsection{Variable Substitution}
With the variable transformation~\eqref{eq.PQ}, 
we note the Jacobian of the integral 
\begin{align}
\frac{\mathrm{d}^3p_k}{p_k^0} \frac{\mathrm{d}^3p_l}{p_l^0}
\frac{\mathrm{d}^3p_i}{p^{0}_i} \frac{\mathrm{d}^3p_j}{p^{ 0}_j}
= \mathrm{d}\mu(P) \mathrm{d}\mu(P') \mathrm{d}\mu(Q) \mathrm{d}\mu(Q')\,,
\end{align}
with
\begin{align}
\mathrm{d}\mu(P) \equiv\;& \mathrm{d}^4P\,\theta(P^0)\,\theta(P^2),\\
\mathrm{d}\mu(Q) \equiv\;& \mathrm{d}^4Q\,\delta(Q^2+P^2) \delta(P\cdot Q),
\end{align}
and we express the transition rate by the differential cross-section
\begin{align}
W_{kl\to ij} = P^2 \sigma_{kl\to ij}(P, -\hat{Q}\cdot\hat{Q'}) \delta^{(4)}(P-P')\,,
\end{align}
where $\hat{V}\equiv V/\sqrt{V\cdot V}$ denotes the direction of a four-vector.
Then, the integrals are re-expressed as
\begin{align}
\mathcal{J}^{(a,b,d,e,f)}_{kl\rightarrow ij}
=\;&	\int \mathrm{d}\mu(P)(P^2)^a e^{-\beta u \cdot P} (u\cdot P)^b
	\int \mathrm{d}\mu(Q)\nn
\\& \times
	\mathrm{d}\mu(Q')(u\cdot Q)^d (u\cdot Q')^e (Q\cdot Q')^f\nn
\\& \times
	\sigma_{kl\to ij}(P,-\hat{Q}\cdot\hat{Q}')\,,\\
\mathcal{J}^{(a,b,e,f)}_{kl \to ij}
=\;&	\int \mathrm{d}\mu(P)(P^2)^a e^{-\beta u \cdot P} (u\cdot P)^b
	\int \mathrm{d}\mu(Q)\nn
\\& \times
    \mathrm{d}\mu(Q')\frac{(u\cdot Q')^e(Q\cdot Q')^f}{u\cdot P+u\cdot Q}\nn
\\& \times
    \sigma_{kl\to ij}(P,-\hat{Q}\cdot\hat{Q}')\,.
\end{align}

As pointed out in \cite{DeGroot:1980dk}, it will be more straight-forward to perform relative-momenta integrals in the center-of-mass frame of the collisions, where we label the transformed velocity as $U^{\mu} = \{U^0, \vec{U}\}$, and
\begin{align}
\begin{split}
&
    P^{\mu} = (P, \vec{0}), \quad
	Q^{\mu} = (0, \vec{Q}),\quad
	Q'^{\mu} = (0, \vec{Q'}),\\
&
    Q\cdot Q' = -P^2 \cos\Theta,\quad
	u\cdot P = U^0 P,\\
& 
	u\cdot Q = -U\, P \cos\theta,\quad
	u\cdot Q' = -U\, P \cos\theta'.
\end{split}
\end{align}
with $\Theta$ being the angle between $\vec{Q}$ and $\vec{Q'}$, while $\theta$($\theta'$) the angle between $\vec{Q}$($\vec{Q'}$) and $\vec{U}$, the delta function ensures that $|\vec{Q}| = |\vec{Q'}| = P$. Particularly, the velocity is measured in the center-of-mass frame of the collisions, they can be expressed by the lab-frame quantities as $U^0 = (u\cdot P)/\sqrt{P^2}$, $U = \sqrt{U_0^2-1}$.

\subsubsection{Relative Momentum Integral}
To compute the integrals, we perform the relative momentum integration ($\int \mathrm{d}\mu(Q) \mathrm{d}\mu(Q')$).
We sequentially apply the following mathematical relations:
\begin{itemize}
\item 
    Partial-wave expansion --- expand the collision kernel over the Legendre polynomials, $\mathrm{P}_l(x)$,
\begin{align}
&	x^d  \sigma_{kl\to ij}(P,x) 
=		\sum_{g=0}^{\infty} \sigma^{(d,g)}_{kl\to ij}(P) \mathrm{P}_g(x)\,,\\
&	\sigma^{(d,g)}_{kl\to ij}(P) 
\equiv	\frac{2g+1}{2}\int_{-1}^{1} \, x^d  \mathrm{P}_g(x) \sigma_{kl\to ij}(P,x) \mathrm{d}x,
\end{align}
\item 
    Additional theorem of the Legendre polynomials,
\begin{align}
	&\int  \mathrm{d}\phi \,\mathrm{d}\phi' \mathrm{P}_g(\cos\Theta)\nn\\
=\;&	(2\pi)^2 \mathrm{P}_{g}(\cos\theta) \mathrm{P}_{g}(\cos\theta')\,,
\end{align}
\item 
    Integral of the Legendre polynomials (cf. Eq.(7.126) of \cite{gradshteyn})
\begin{align}
	\int_{-1}^{1}   x^e \mathrm{P}_{g}(x)\,\mathrm{d}x
=\left\{\begin{array}{ll}
	\frac{(1+(-1)^{e-g})\,  e!}{(e-g)!! (e+g+1)!!}, & g\leq e,\\
	~\\
	0,	&g > e. \\
\end{array}\right.
\end{align}
when $z\in \mathbb{R}$ and $z>1$, 
\begin{align}
	\int_{-1}^{1}  \frac{\mathrm{P}_{g}(x)}{z-x}\,\mathrm{d}x
=&	\sum_{m=0}^{\infty} z^{-m-1} \int_{-1}^{1} x^m \mathrm{P}_{g}(x)\,\mathrm{d}x
\nn\\
=&	\sum_{n=0}^{\infty} 2 z^{-2n-g-1} \frac{(g+2n)!}{(2n)!! (2g+2n+1)!!}.
\end{align}
For later convenience, we use the shorthand that
\begin{align}
C_{e,g} \equiv \frac{1+(-1)^{e-g}}{2}\frac{e!}{(e-g)!! (e+g+1)!!}\,.
\end{align}
\end{itemize}

With these, we can integrate out the relative momenta, $Q$ and $Q'$, and reach the following relations respective for five- and four-indices $\mathcal{J}$ integral
\begin{align}
&	\int \mathrm{d}\mu(Q) \mathrm{d}\mu(Q')
 	(u\cdot Q)^d (u\cdot Q')^e (Q\cdot Q')^f\nn\\
&\qquad\times
    \sigma_{kl \to ij}(P,-\hat{Q}\cdot\hat{Q}') \nn\\
=\;&	(-1)^{f} 4\pi^2 \sum_{g=0}^{\min(d,e)}
	C_{d,g} C_{e,g}
	P^{d+e+2f} U^{d+e} \sigma^{(f,g)}_{kl\to ij}(P)\,,
\end{align}
and
\begin{align}
&	\int \mathrm{d}\mu(Q) \mathrm{d}\mu(Q')
 	\frac{(u\cdot Q')^e (Q\cdot Q')^f}{u\cdot P+u\cdot Q}\sigma_{kl \to ij}(P,-\hat{Q}\cdot\hat{Q}') \nn\\
=\;&	\sum_{h=0}^{[e/2]} 2\pi^2 (-1)^{e+f} C_{e,e-2h}
	P^{e+2f-1} U^e  \sigma^{(f,e-2h)}_{kl\to ij}(P)
\nn\\&\quad\times
	\int_{-1}^{1} \mathrm{d}\cos\theta \frac{ \mathrm{P}_{e-2h}(\cos\theta) }{U^0 - U \cos\theta}\,.
\end{align}

\subsubsection{Total Momentum Integral}
Finally, the integrals over the total-momentum would be highly simplified in the fluid co-moving frame, $u=(1,\vec{0})$, and with the variable substitution that $P^2=\beta^{-2} v^2$, $P_0=\beta^{-1}v\cosh\xi$, so that
$\mathrm{d}\mu(P)=\beta^{-4} v^3\mathrm{d}v\,\sinh^2\xi\mathrm{d}\xi\,\mathrm{d}\Omega_P$

Following steps will be taken in order:
\begin{itemize}
\item 
    Variable substitution: $P=\beta^{-1} v$, $u\cdot P = \beta^{-1}v\cosh\xi$, $U^0 = (u\cdot P)/\sqrt{P^2} = \cosh\xi$, $U = \sqrt{U_0^2-1} = \sinh\xi$,
\item 
    The trick that $\cosh^b\xi \,e^{-v\cosh\xi} = (-1)^b\frac{\partial^b}{\partial v^b} e^{-v\cosh\xi}$,
\item 
    Integral expression of modified Bessel function of the second kind (see Eq.(10.32.8) of \cite{nist}):
\begin{align}
	\int_0^{\infty} \mathrm{d}\xi\, e^{-z \cosh\xi} \sinh^{2n}\xi
=	(2n-1)!! \frac{K_n(z)}{z^n}\,,
\end{align}
\item 
    The following integral (see App.~\ref{sec.eq.intgral_F} for proof)
\begin{align}
\begin{split}
&	\int_{-1}^{1} \mathrm{P}_{\ell}(x) \mathrm{d}x 
	\int_0^{\infty}\mathrm{d}\xi\, 
	\frac{ \sinh^{\ell+2h+2}\xi}{\cosh\xi - x\sinh\xi} e^{-v\cosh\xi}
\\=\;&
	2 (\ell!)  \big( \partial_v^2 -1 \big)^h\frac{K_0(v)}{v^{\ell+1}} \,.
\end{split}\label{eq.intgral_F}
\end{align}
\end{itemize}

Eventually, we have simplify the integrals into one-dimensional ones, which will be performed numerically in practical calculations,
\begin{align}
&	\mathcal{J}^{(a,b,d,e,f)}_{kl\rightarrow ij} \nn\\
=\;&	(-1)^{f} 4\pi^2 \sum_{g=0}^{\min(d,e)} C_{d,g} C_{e,g}
	\int \mathrm{d}\mu(P) e^{-\beta u \cdot P} (u\cdot P)^b 
\nn\\&\quad\times
    P^{2a+d+e+2f} U^{d+e} \sigma^{(f,g)}_{kl\to ij}(P) \nn\\
\begin{split}
=\;&	\frac{(-1)^{f} 16\pi^3 (d+e+1)!!}{\beta^{2a+b+d+e+2f+4}}
	\sum_{g=0}^{\min(d,e)} C_{d,g} C_{e,g}
\\&\quad\times
	\int_0^{\infty} \mathrm{d}v\, v^{2a+b+d+e+2f+3}  \sigma^{(f,g)}_{kl\to ij}(v/\beta) 
\\&\quad\times
	(-1)^b \frac{\partial^b}{\partial v^b} \Big[v^{-\frac{d+e}{2}-1}K_{\frac{d+e}{2}+1}(v)\Big] \,,
\end{split}\label{eq.J5_result}
\end{align}
and
\begin{align}
\begin{split}
\mathcal{J}^{(a,b,e,f)}_{kl \to ij}
=\;&	\sum_{h=0}^{[e/2]} 2\pi^2 (-1)^{e+f} C_{e,e-2h}
	\int \mathrm{d}\mu(P) e^{-\beta u \cdot P}
\\&\quad\times
	(u\cdot P)^bP^{2a+e+2f-1} U^e  \sigma^{(f,e-2h)}_{kl\to ij}(P)
\\&\quad\times
	\int_{-1}^{1} \mathrm{d}\cos\theta \frac{ \mathrm{P}_{e-2h}(\cos\theta) }{U^0 - U \cos\theta}
\end{split}\nn\\
\begin{split}
=\;&	 \frac{(-1)^{b+e+f} 16\pi^3}{\beta^{2a+b+e+2f+3}} \sum_{h=0}^{[e/2]} (e-2h)! C_{e,e-2h}
\\&\quad\times
	\int_0^{\infty} \mathrm{d}v\, v^{2a+b+e+2f+2}  \sigma^{(f,e-2h)}_{kl\to ij}(\frac{v}{\beta}) \,
\\&\quad\times
	\sum_{g=0}^{h} (-1)^{h-g}\frac{\partial^{b+2g}}{\partial v^{b+2g}} \frac{K_0(v)}{v^{e-2h+1}}\,.
\end{split}\label{eq.J4_result}
\end{align}

\subsubsection{Proof of Eq.~\eqref{eq.intgral_F}}\label{sec.eq.intgral_F}
In this subsubsection, we provide a proof of the relation outlined in Eq.~\eqref{eq.intgral_F}.
We denote that
\begin{align}
F_{\ell,h}(v) \equiv&
	\int_{-1}^{1} \mathrm{P}_{\ell}(x) \mathrm{d}x 
	\int_0^{\infty}\mathrm{d}\xi\, 
	\frac{e^{-v\cosh\xi} \sinh^{\ell+2h+2}\xi}{\cosh\xi - x\sinh\xi}
	,
\end{align}
and one can find the iteration relations
\begin{align}
F_{\ell+1,h}(v) \equiv\;&
	\int_{-1}^{1} \mathrm{P}_{\ell+1}(x) \mathrm{d}x 
	\int_0^{\infty}\mathrm{d}\xi\, 
	\frac{ \sinh^{\ell+1+2h+2}\xi}{\cosh\xi - x\sinh\xi} \nn
\\&\quad\times
	e^{-v\cosh\xi}\nn\\
=\;&
	- \frac{2\ell+1}{\ell+1} \partial_v F_{\ell,h}(v)
	-\frac{\ell}{\ell+1} \Big(\partial_v^2 F_{\ell-1,h}(v)\nn
\\&\,
	- F_{\ell-1,h}(v)\Big)
	 -2\delta_{\ell,0} (2h+1)!! \frac{K_{h+1}(v)}{v^{h+1}}
\end{align}

\begin{align}
F_{\ell,h+1}(v) =\;& \partial_v^2 F_{\ell,h}(v)  - F_{\ell,h}(v) \,,\\
\begin{split}
F_{\ell+1,h}(v) =\;& 
	- \frac{2\ell+1}{\ell+1} \partial_v F_{\ell,h}(v)
\\&-\frac{\ell}{\ell+1}
	\Big(\partial_v^2 F_{\ell-1,h}(v)  - F_{\ell-1,h}(v)\Big)
\\&
	 - 2\delta_{\ell,0} (2h+1)!! \frac{K_{h+1}(v)}{v^{h+1}} \,.
\end{split}
\end{align}
Noting that $F_{0,0}(v) = 2 v^{-1} K_0(v)$ can be computed directly, therefore,
\begin{align}
&F_{\ell,0}(v) =\; 2 (\ell!) \frac{K_0(v)}{v^{\ell+1}} \,,\\
&F_{\ell, h}(v) = 2 (\ell!)  \big( \partial_v^2 -1 \big)^h\frac{K_0(v)}{v^{\ell+1}} \,.
\end{align}

\section{hard-sphere Potential}
\label{app.hard_sphere}

In this Appendix, we take the hard-sphere potential,
\begin{align}
\sigma_{kl\to ij}(P) =\;& \sigma_{kl\to ij}\,,\\
\sigma^{(d,g)}_{kl\to ij}(P) =\;&	(2g+1)  C_{d,g} \,\sigma_{kl\to ij},
\end{align}
and evaluate the $\mathcal{J}$ and $\mathcal{K}$ integrals, which will be used to compute the collision matrix and the transport coefficients. To compute the five-index  $\mathcal{J}$ integrals, following steps will be performed in order:
\begin{itemize}
\item 
    Integration by parts --- for function $f(v)$ satisfying certain boundary conditions,
\begin{align}
\begin{split}
&
    \int_0^\infty v^n  (-1)^b \frac{\partial^b f(v)}{\partial v^b} \mathrm{d}v 
\\=\;&
    \frac{n!}{(n-b)!} \int_0^\infty v^{n-b} f(v) \mathrm{d}v\,,
\end{split}
\end{align}
\item 
    Eq.~(10.43.19) of Ref.~\cite{nist}, i.e.,
\begin{align}
\begin{split}
&
    \int_0^{\infty} x^{\mu} K_\nu(x) \mathrm{d}x 
\\=\;&
    2^{\mu-1} \Gamma((\mu+1+\nu)/2) \Gamma((\mu+1-\nu)/2)\,.
\end{split}
\end{align}
\end{itemize}

We find
\begin{align}
\begin{split}
&    \mathcal{J}^{(a,b,d,e,f)}_{kl\rightarrow ij}
\\=\;&
    (-1)^{f} \frac{16\pi^3 (d+e+1)!!}{\beta^{2a+b+d+e+2f+4}} 
    2^{a+f} \sigma_{kl\to ij} 
\\&\times
	\frac{(2a+b+d+e+2f+3)!}{(2a+d+e+2f+3)!!}
	(a+f)!
\\&\times
	\sum_{g=0}^{\min(d,e,f)} (2g+1) C_{d,g} C_{e,g}C_{f,g} 
	\,,
\end{split}
\end{align}
and
\begin{align}
\begin{split}
&	\mathcal{J}^{(a,b,e,f)}_{kl \to ij}
\\=\;&
    \frac{(-1)^{e+f} 16\pi^3}{\beta^{2a+b+e+2f+3}}
    \sigma_{kl\to ij} 
    \sum_{h=0}^{[e/2]}\sum_{g=0}^{h}\Bigg[ (-1)^{h-g} 
\\&\times
   (e-2h)!(2e-4h+1) C_{e,e-2h}C_{f,e-2h}
\\&\times
    \frac{(2a+b+e+2f+2)!}{(2a+e+2f+2-2g)!}
    2^{2a+2f-2g+2h}
\\&\times
    ((a+f-g+h)!)^2\bigg].	
\end{split}
\end{align}

\section{Collision Integrals for Leading-Order pQCD-cross sections}
\label{sec.pQCD_cross}
In this appendix, we compute collisions integrals taking leading order pQCD cross-sections in consideration.
For completeness, we list the differential cross-sections for different processes as follows.
\begin{align}
\begin{split}
    \sigma_{gg\rightarrow gg}
=\;& 
    \frac{9\alpha_s^2}{8s} \bigg[3 - \frac{tu}{(s+m_D^2)^2} - \frac{su}{(t-m_D^2)^2} 
\\&\quad
    - \frac{st}{(u-m_D^2)^2} \bigg]
    \,,
\end{split}\label{eq.pQCD_1}
\end{align}
\begin{align}
\begin{split}
    \sigma_{qg \rightarrow qg}
=\;& 
    \frac{\alpha_s^2}{4s} \bigg[-\frac{4}{9}\frac{su}{(s+m_q^2)^2} -\frac{4}{9}\frac{su}{(u-m_q^2)^2} 
\\&\quad
    - \frac{2su}{(t-m_D^2)^2} 
    - \frac{su}{(t-m_D^2)(s+m_q^2)} 
\\&\quad
    - \frac{su}{(t-m_D^2)(u-m_q^2)} \bigg]\,,
\end{split}
\end{align}
\begin{align}
\begin{split}
    \sigma_{qq' \to qq' }
=\;&	
    \frac{\alpha_s^2}{9s} \frac{s^2+u^2}{(t-m_D^2)^2}\,,
\end{split}
\end{align}
\begin{align}
\begin{split}
\sigma_{qq\rightarrow qq}
=\;& 
    \frac{\alpha_s^2}{9s} \bigg[\frac{s^2+u^2}{(t-m_D^2)^2} + \frac{s^2+t^2}{(u-m_D^2)^2}
\\&\quad
    - \frac{2}{3}\frac{s^2}{(u-m_D^2)(t-m_D^2)} \bigg]\,,
\end{split}
\end{align}
\begin{align}
\begin{split}
\sigma_{q\bar{q} \to q\bar{q}}
=\;&
    \frac{\alpha_s^2}{9s}  \bigg[\frac{s^2+u^2}{(t-m_D^2)^2} 
    + \frac{t^2+u^2}{(s+m_D^2)^2}
\\&\quad
    - \frac{2}{3}\frac{u^2}{(s+m_D^2)(t-m_D^2)} \bigg]\,,
\end{split}
\end{align}
\begin{align}
\begin{split}
\sigma_{q\bar{q} \to q'\bar{q}'}
=\;&	
    \frac{\alpha_s^2}{9s} \frac{t^2+u^2}{(s+m_D^2)^2} \,,
\end{split}
\end{align}
\begin{align}
\begin{split}
\sigma_{q\bar{q} \to gg} 
=\;&
    \frac{2\alpha_s^2}{3s}\bigg[\frac{4}{9}\frac{tu}{(t-m_q^2)^2} + \frac{4}{9}\frac{tu}{(u-m_q^2)^2} \\
\;& 
    +2\frac{tu}{(s+m_D^2)^2} 
    +\frac{tu}{(s+m_D^2)(t-m_q^2)}\\
\;& 
    +\frac{tu}{(s+m_D^2)(u-m_q^2)}\bigg]\,,
\end{split}
\end{align}
with following relations due to symmetry,
\begin{align}
\sigma_{gg \to q\bar{q}} =\;& 
    \frac{9}{64} \sigma_{q\bar{q} \to gg}\,, \\
\sigma_{\bar{q}\bar{q}' \to \bar{q}\bar{q}'} =\;&
    \sigma_{q\bar{q}' \to q\bar{q}'} = \sigma_{qq' \to qq'}\,, \\
\sigma_{\bar{q}g \rightarrow \bar{q}g} =\;&
    \sigma_{qg \rightarrow qg}\,,\\
\sigma_{\bar{q}\bar{q}\rightarrow \bar{q}\bar{q}} =\;&
    \sigma_{qq\rightarrow qq}\,.
    \label{eq.pQCD_11}
\end{align}
Here screen masses $m_D$ and $m_q$ are introduced for gluons and quarks to avoid infrared divergence. In detailed calculation, they are non-trivial dependent on temperature and chemical potentials, and the dependence relies on the details of interaction included~\cite{Xu:2004mz},
\begin{align}
&m_D=\sqrt{\frac{8\alpha_s}{\pi}(N_c+N_f)}\,T\,,\\
&m_q=\sqrt{\frac{2(N_c^2-1)\alpha_s}{\pi N_c}}\,T\,.
\end{align}
with number of color $N_c=3$ and number of flavor $N_f=2$, and the coupling constant is evaluated at the thermal energy scale $\alpha_s = \alpha_s(\mu^2_R=4\pi^2T^2)$. The running of the coupling is driven by
\begin{align}
\mu^2_R\frac{d\alpha_s(\mu^2_R)}{d\mu^2_R}=-(b_0\alpha_s^2+b_1\alpha_s^3+b_2\alpha_s^4),
\end{align}
with fixed point $\alpha_s(\mu_R^2 = 25 \,\text{GeV}^2)=0.2034$ and coefficients $b_0=\frac{33-2N_f}{12\pi}$,  $b_1=\frac{153-19N_f}{24\pi^2}$ and $b_2=\frac{2857-\frac{5033}{9}N_f+\frac{325}{27}N_f^2}{128\pi}$.

We express the Mandelstam variables as
\begin{align}
s = v^2, \qquad
t = -\frac{v^2}{2}(1-x),\qquad
u = -\frac{v^2}{2}(1+x),
\end{align}
and evaluate the $\mathcal{J}$ and $\mathcal{K}$ integrals, and compute the collision matrix as well as the transport coefficients.
Performing integral by part and expanding the Legendre polynomial, we express the five-index and four-index $\mathcal{J}$ integrals as
\begin{align}
\begin{split}
&   \mathcal{J}^{(a,b,d,e,f)}_{kl\rightarrow ij}\\
=\;&
	(-1)^{f} \frac{16\pi^3 (d+e+1)!!}{\beta^{2a+b+d+e+2f+4}}
	\frac{\alpha^2_s}{2} \sum_{g=0}^{\min(d,e)}
	\Bigg[ 
	\\&
	C_{d,g} C_{e,g} \frac{2g+1}{2^{g+1}}
	\sum_{h=0}^{[\frac{g}{2}]}\bigg[
	\frac{(-1)^h(2g-2h)!}{h!(g-h)!(g-2h)!}
	\\&\times\sum_{m=0}^{[\frac{b}{2}]}\bigg(
	\frac{(-1)^{m}\,(2m-1)!!\,b!}{(2m)!(b-2m)!}
	\\&\times
    \mathcal{I}^{(\frac{d+e}{2}+1+b-m,a+f+1,f+g-2h)}_{kl\to ij}\bigg)\bigg]\Bigg]\,,
	\end{split}
\end{align}

and
\begin{align}
\begin{split}
&	\mathcal{J}^{(a,b,e,f)}_{kl \to ij}\\
=\;&
    \frac{(-1)^{e+f} 16\pi^3}{\beta^{2a+b+e+2f+3}}
    \frac{\alpha^2_s}{2}
    \sum_{h=0}^{[e/2]}\sum_{g=0}^{h}\sum_{m=0}^{b+2g}\sum_{n=0}^{[\frac{b+2g-m}{2}]}\Bigg[
\\&
    (-1)^{n+h+g}
    \frac{2e-4h+1}{2^{e-2h+1}}	
    C_{e,e-2h}
\\&\times
    \frac{(b+2g)!\,(2n-1)!!\,(e-2h+m)!}{m!\,(2n)!\,(b+2g-m-2n)!}
\\&\times \sum_{j=0}^{[\frac{e-2h}{2}]}\bigg[
    \frac{(-1)^j(2e-4h-2j)!}{j!(e-2h-j)!(e-2h-2j)!} 
\\&\times
     \mathcal{I}^{(b+2g-m-n,a+f+h-g+1,f+e-2h-2j)}_{kl\to ij}\bigg]\Bigg]\,,
\end{split}
\end{align}
where
\begin{align}
\begin{split}
    \mathcal{I}^{(\nu,n,\ell)}_{ab\to cd} 
\equiv\;& 
    \frac{2}{\alpha_s^2} \int_{-1}^{1} x^{\ell} \mathrm{d}x
    \int_0^\infty \mathrm{d}v\,
\\&\times
     v^{\nu+2n-1} K_\nu(v) \sigma_{ab\to cd}(v, x).
\end{split}
\label{eq.I_integral}
\end{align}
General result of such integration terms will be given in the succeeding subsection.

\subsection{calculation of $\mathcal{I}^{(\nu,n,l)}$}
\label{Inl}
We note that the express of differential cross-sections contain many repeated pieces. Hence, to simplify the calculation, we define the following integrals which will be used later,
\begin{align}
\begin{split}
&    \mathbf{A}_\nu^{(n)} (z) 
\\\equiv\;& 
    \int_0^\infty \mathrm{d}x \,
    \frac{x^{\nu+2n+1} K_{\nu}(x)}{x^2+z}
\\=\;&
    (-z)^n \mathbf{A}_{\nu}^{(0)}(z)
    + \sum_{k=0}^{n-1}\Big[ 2^{\nu+2n-2-2k}
\\&\times
    \Gamma (\nu+n-k) \,\Gamma (n-k) \,(-z)^k\Big]\,,
\end{split}
\end{align}
\begin{align}
\begin{split}
&    \mathbf{B}_\nu^{(n)} (z) 
\\\equiv\;& 
    - \frac{\mathrm{d}}{\mathrm{d}z} \mathbf{A}_\nu^{(n)} (z)
\\=\;&
    (-z)^{n-1} \frac{2(\nu+n) \mathbf{A}_{\nu}^{(0)}(z) - \mathbf{A}_{\nu+1}^{(0)}(z)}{2}
\\&
    +\sum_{k=1}^{n-1} \Big[ k\,2^{\nu+2n-2-2k} \, \Gamma (\nu+n-k)
\\&\quad\times
    \Gamma (n-k) \,(-z)^{k-1} \Big]\,,
\end{split}
\end{align}
\begin{align}
\begin{split}
&\mathbf{C}_\nu^{(n,m)} (z) 
\\\equiv\;& 
    \frac{1}{z}\int_1^{\infty} \frac{\mathrm{d}t}{t^{m+1}} \mathbf{A}_\nu^{(n)} (t\,z)\,,
\end{split}
\end{align}
\begin{align}
\begin{split}
&   \mathbf{T}_\nu^{(n,m)} (y,z)
\\\equiv\;& \int_1^{\infty} \frac{t^{-m}\mathrm{d}t}{t\,z-y} \Big(\mathbf{A}_\nu^{(n)} (t\,z) - \mathbf{A}_\nu^{(n)} (y)\Big)\,,
\end{split}
\end{align}
for $z>0$, $\nu\geq0$, $n\geq0$.

where
$y\equiv \frac{2}{1-x}$, $x = (1-2/y)$, $(1+x) = 2(1-1/y)$, $\mathrm{d}x = 2\mathrm{d}y/y^2$.
Then, we can calculate the $\mathcal{I}^{(\nu,n,l)}$ integrals, defined in Eq.~\eqref{eq.I_integral}, for all channels:

\begin{align}
    \mathcal{I}^{(\nu,n,l)}_{q\bar{q}\to q'\bar{q}'} 
=\;&
    \frac{2(1+(-1)^l)(2+l)}{9(1+l)(3+l)}\mathbf{B}_\nu^{(n)}(m_D^2)\,,
\end{align}

\begin{align}
\begin{split}
&    \mathcal{I}^{(\nu,n,l)}_{qq'\to qq'}
\\=\;&
    \frac{4(-1)^l}{9\,m_D^2}\mathbf{A}_\nu^{(n)}(m_D^2) -\frac{8}{9}\sum_k\frac{(-2)^{l-k}l!}{k!(l-k)!}\times
\\&
    \Bigg[(l-k)\mathbf{C}_\nu^{(n,l-k)}(m_D^2)
    -(l-k+1)\mathbf{C}_\nu^{(n,l-k+1)}(m_D^2)
\\&
    +\frac{l-k+2}{2}\mathbf{C}_\nu^{(n,l-k+2)}(m_D^2)\Bigg],
\end{split}
\end{align}

\begin{align}
\begin{split}
&    \mathcal{I}^{(\nu,n,l)}_{qq\to qq}
\\=\;&
    (1+(-1)^l)\Bigg[
    \mathcal{I}^{(\nu,n,l)}_{qq'\to qq'}
    +\frac{8\,\delta_{l0}}{27}\mathbf{T}_\nu^{(n,1)}(2m_D^2,m_D^2)
\\&
    +\frac{8(1-\delta_{l0})}{27}\sum_k\frac{(-2)^{l-1-k} (l-1)!}{k!(l-1-k)!}\mathbf{C}_\nu^{(n,l-k)}(m_D^2)
    \Bigg]\,,
\end{split}
\end{align}

\begin{align}
\begin{split}
&    \mathcal{I}^{(\nu,n,l)}_{q\bar{q}\to q\bar{q}} 
\\=\;&
    \frac{4}{9\,m_D^2}\Big((-1)^l+\frac{(-1)^l+2l+3}{6(1+l)(2+l)}\Big)
    \mathbf{A}_\nu^{(n)}(m_D^2)
\\&
    +\frac{2(1+(-1)^l)(2+l)}{9(1+l)(3+l)}\mathbf{B}_\nu^{(n)}(m_D^2)
\\&
    -\frac{8}{9}\sum_k\frac{(-2)^{l-k}l!}{k!(l-k)!}
    \Bigg[
    (l-k)\mathbf{C}_\nu^{(n,l-k)}(m_D^2)
\\&
    -(l-k+\frac{2}{3})\mathbf{C}_\nu^{(n,l-k+1)}(m_D^2)
\\&
    +(\frac{l-k}{2}+\frac{2}{3})\mathbf{C}_\nu^{(n,l-k+2)}(m_D^2)
\Bigg]\,,
\end{split}
\end{align}

\begin{align}
\begin{split}
&    \mathcal{I}^{(\nu,n,l)}_{gg\to gg} 
\\=\;&
    \frac{9(1+(-1)^l)}{l+1}\Bigg[
    2^{2n+\nu-6} 3 \,\Gamma(n-1)\Gamma(n+\nu-1)
\\&
    -\frac{\mathbf{B}_\nu^{(n)} (m_D^2)}{8(3+l)}
    +\frac{l+1}{2}\sum_k\frac{(-2)^{l-k} l!}{k!(l-k-1)!}
\\&
    \bigg(\frac{l-k+1}{l-k}\mathbf{C}_\nu^{(n,l-k+1)}(m_D^2)
    -\mathbf{C}_\nu^{(n,l-k)}(m_D^2)
\bigg)\Bigg]\,,
\end{split}
\end{align}

\begin{align}
\begin{split}
&    \mathcal{I}^{(\nu,n,l)}_{q\bar{q}\to gg}
\\=\;&
    \frac{4(1+(-1)^l)}{3(3+l)(1+l)}\mathbf{B}_\nu^{(n)} (m_D^2)
\\&
    +(1+(-1)^l)\frac{8}{3}\sum_k\frac{(-2)^{l-k} l!}{k!(l-k)!} 
    \times
\\&
    \Bigg[-\frac{4(l-k+1)}{9}\mathbf{C}_\nu^{(n,l-k+1)} (m_q^2)
\\&
    +\frac{4(l-k+2)}{9}\mathbf{C}_\nu^{(n,l-k+2)}(m_q^2)
\\&
    +\mathbf{T}_\nu^{(n,l-k+2)} (m_D^2,m_q^2)
    -\mathbf{T}_\nu^{(n,l-k+3)} (m_D^2,m_q^2)\Bigg]\,,
\end{split}
\end{align}
and
\begin{align}
\begin{split}
&   \mathcal{I}^{(\nu,n,l)}_{qg\to qg}
\\=\;&
    \frac{(-1)^l+2l+3}{9(1+l)(2+l)} \mathbf{B}_\nu^{(n)} (m_q^2)
\\&
    +\sum_k\frac{(-2)^{l-k} l!}{k!(l-k)!} \Bigg[\frac{4(-1)^l}{9\,m_q^2} \mathbf{A}_\nu^{(n)} (m_q^2)
\\&
    -\frac{4(-1)^l(l-k+1)}{9} \mathbf{C}_\nu^{(n,l-k+1)} (m_q^2)
\\&
    -2(l-k)\mathbf{C}_\nu^{(n,l-k)} (m_D^2) 
\\&
    +2(l-k+1) \mathbf{C}_\nu^{(n,l-k+1)} (m_D^2)
\\&
    +\mathbf{T}_\nu^{(n,l-k+1)} (m_q^2,m_D^2)
\\&
    -\mathbf{T}_\nu^{(n,l-k+2)} (m_q^2,m_D^2)
\\&
    -\mathbf{T}_\nu^{(n,l-k+1)} (m_D^2+m_q^2,m_D^2) 
\\&
    +\mathbf{T}_\nu^{(n,l-k+2)} (m_D^2+m_q^2,m_D^2)
\\&
    - (-1)^l\mathbf{T}_\nu^{(n,l-k+2)} (m_D^2+m_q^2,m_q^2)
\Bigg]\,.
\end{split}
\end{align}

One may wonder that computing those newly defined functions with different $\nu$, $n$, and $m$ indices might also be computational expansive. Here, we derive their recurrence relations so that one can compute a small number of functions and then generate others accordingly.
We note that $t^{-1}(t\,z-y)^{-1} = (z/y) ((t\,z-y)^{-1}-(t\,z)^{-1})$, exploiting the recurrence relations of modified Bessel functions $K_\nu(x)$, the following useful recurrence relations are obtained.
\begin{align}
\mathbf{T}_\nu^{(n,m+1)} (y,z) =& \frac{z}{y} \mathbf{T}_\nu^{(n,m)} (y,z) - \frac{z}{y}\mathbf{C}_\nu^{(n,m)} (z) \nn
\\&
    +\frac{1}{m\,y}\mathbf{A}_\nu^{(n)} (y),\\
    \mathbf{T}_{\nu+1}^{(n,m)} (y,z)
=\;&
    2\nu\mathbf{T}_{\nu}^{(n,m)} (y,z)+\mathbf{T}_{\nu-1}^{(n+1,m)} (y,z),\\
    \mathbf{A}_{\nu+1}^{(n)} (z)
=\;&
    2\nu\mathbf{A}_{\nu}^{(n)} (z)+\mathbf{A}_{\nu-1}^{(n+1)} (z),\\
    \mathbf{A}_{\nu}^{(n+1)} (z)
=\;&
    -z\mathbf{A}_{\nu}^{(n)} (z)\nn
\\&
    +2^{\nu+2n}\Gamma(\nu+n+1)\Gamma(n+1),\\
    \mathbf{C}_{\nu+1}^{(n,m)} (z) 
=\;&
    2\nu\mathbf{C}_{\nu}^{(n,m)} (z)+\mathbf{C}_{\nu-1}^{(n+1,m)} (z).\\
\mathbf{C}_{\nu}^{(n,m)}(z)=&\,\frac{2^{\nu+2n-2}\Gamma(\nu+n)\Gamma(n)}{mz}\nn
\\&
    -z\,\mathbf{C}_{\nu}^{(n-1,m-1)} (z).
\end{align}
It shall be useful to note the asymptotic behavior of $\mathbf{A}_\nu^{(n)} (z)$ at large $z$ limit, 
\begin{align}
\mathbf{A}_\nu^{(n)} (z) = & 
\frac{1}{z} \sum_{k=0}^{N_k} \frac{(-1)^k}{z^k} \int_0^\infty \mathrm{d}x \,
x^{\nu+2n+2k+1} K_{\nu}(x)\nn\\
=&	\frac{1}{z} \sum_{k=0}^{N_k} \frac{(-1)^k}{z^k} 
2^{\nu+2n+2k} \Gamma(\nu+n+k+1) \nn\\
&\quad\times\,
\Gamma(n+k+1)\,.
\end{align}
\end{appendix}
\clearpage
\end{document}